\documentclass[epsf]{emulateapj}
\shorttitle{M87 UCDs}
\shortauthors{Brodie \etal~}
\def\etal{{\it et al.}}
\def\kms{\,km~s$^{-1}$}

\def\rh{$r_{\rm h}$}
\def\rt{$r_{\rm t}$}
\usepackage{subfigure}

\begin{document}

\title{The Relationships Among Compact Stellar Systems: \\ 
A Fresh View of Ultra Compact Dwarfs}

\author{Jean P.~Brodie\altaffilmark{1}, Aaron J.~Romanowsky\altaffilmark{1}, Jay Strader\altaffilmark{2}, Duncan A.~Forbes\altaffilmark{3}}
\email{brodie@ucolick.org}

\affil{
\altaffilmark{1}UCO/Lick Observatory, University of California, Santa Cruz, CA 95064, USA\\
\altaffilmark{2}Harvard-Smithsonian Center for Astrophysics, Cambridge, MA 02138, USA\\
\altaffilmark{3}Centre for Astrophysics \& Supercomputing, Swinburne University, Hawthorn, VIC 3122, Australia\\
}

\begin{abstract}
We use a combined imaging and spectroscopic survey of the nearby central cluster galaxy, M87, to assemble a sample of 34 confirmed ultra compact dwarfs (UCDs) with half-light radii of
$\ga$~10~pc measured from \emph{Hubble Space Telescope} images. This doubles the existing sample in M87, making it the largest such sample for any galaxy, while
extending the detection of UCDs to unprecedentedly low luminosities ($M_V=-9$). With this expanded sample, we find no correlation between size and luminosity, in contrast to previous suggestions, and no general correlation between size and galactocentric distance.

We explore the relationships between UCDs, less luminous extended clusters (including faint fuzzies),
globular clusters (GCs), as well as early-type galaxies and their nuclei,
assembling an extensive new catalog of sizes and luminosities for stellar systems.
Most of the M87 UCDs follow a tight color-magnitude relation, offset from
the metal-poor GCs. This, along with kinematical differences, 
demonstrates that most UCDs are a distinct population from normal GCs, and
not simply a continuation to larger sizes and higher luminosities.
The UCD color-magnitude trend couples closely with that for Virgo dwarf elliptical nuclei. 
We conclude that the M87 UCDs are predominantly stripped nuclei. The brightest and reddest UCDs may be the remnant nuclei of more massive galaxies while a subset of the faintest UCDs may be tidally limited and related to more compact star clusters. In the broader context of galaxy assembly, blue UCDs may trace halo build-up by accretion of low-mass satellites, while red UCDs may be markers of metal-rich bulge formation in larger galaxies.

\end{abstract}


\section{Introduction}

There is ongoing debate about the existence of a genuine boundary between massive star clusters and compact galaxies (e.g., \citealt{2007ApJ...663..948G,2011PASA...28...77F}). In the decade or so since the discovery \citep{1999A&AS..134...75H,2000PASA...17..227D} and naming \citep{2001ApJ...560..201P} of Ultra Compact Dwarfs (UCDs), numerous
observational studies have sought to understand where these curious objects
fit in the multi-dimensional parameter space for hot (dispersion-supported) stellar systems that encompass massive and dwarf galaxies as well as compact and extended star clusters \citep{2008MNRAS.386..864D,2008MNRAS.389.1924F}.

The UCDs studied to date have typical luminosities of $L \sim 10^7 L_\odot$ and projected half-light
radii of \rh~$\sim$~20 pc.  Such high luminosities and large sizes differ dramatically from classical globular clusters (GCs) with \rh~$\sim$~2--3~pc and $L \la 10^6 L_\odot$. UCDs were first identified in galaxy clusters but have since also been found in low-density environments (e.g., \citealt{2009MNRAS.394L..97H,2011MNRAS.414..739N}).

The formation mechanism for UCDs remains unclear.
One possibility is that some dwarf galaxies in dense environments become stripped down to
their nuclei, which then appear either as normal GCs or as objects of intermediate size
(e.g., \citealt{1994ApJ...431..634B,2001ApJ...552L.105B}). 

Alternatively, UCDs may be a species of GC whose large sizes are the result of star cluster mergers (\citealt{1998MNRAS.300..200K,2002MNRAS.330..642F,2006A&A...448.1031K}),
or of a change in cluster formation physics at high masses \citep{2009ApJ...691..946M}.

An alternative term for UCDs, ``dwarf-globular transition objects'' (coined by \citealt{2005ApJ...627..203H}),
underlines the uncertainty in their identification as either the most massive star clusters or the least massive compact galaxies.
In fact,  there are
indications that no single formation mechanism is responsible for all UCDs
(e.g., \citealt{2006AJ....131.2442M,2009gcgg.book...51H,2010ApJ...712.1191T,2011MNRAS.414..739N,2011A&A...525A..86D,2011MNRAS.412.1627C}).
The challenge is to discover which subsamples of UCDs correspond to which
formational channels and to understand the determining factors (mass, environment, orbit, etc).

The stripped dwarf galaxy scenario has received support from several pieces of observational evidence, including mass-to-light ratios, and size-luminosity and color-magnitude trends.
In some (but not all) cases, UCDs apparently have elevated dynamical mass-to-light
ratios that could imply non-baryonic dark matter (e.g., \citealt{2005ApJ...627..203H,2008MNRAS.391..942B}). 
There have been reports of a strong size-luminosity correlation that differs dramatically
from the nearly constant size of classical GCs
(e.g., \citealt{2007A&A...469..147R,2008AJ....136..461E,2008MNRAS.386..864D,2011MNRAS.414..739N,2011ApJ...737L..13M}).
This trend seemed to bear a rough resemblance to the size-luminosity relations of
dwarf elliptical (dE) nuclei (e.g., \citealt{2006ApJS..165...57C}).

Other similarities were noted between UCDs and nuclei in color-magnitude space
\citep{2006ApJS..165...57C,2008AJ....136..461E,2011MNRAS.414..739N}.
However, the limited range of UCD luminosities in these studies made such comparisons
difficult to interpret and, as we report in this paper, re-evaluation is required once the sample is expanded to
include fainter UCDs.

In parallel to the expanding recognition of UCDs, 
a menagerie of very faint, extended clusters have been discovered around various galaxies
(e.g., \citealt{2000AJ....120.2938L,2005MNRAS.360.1007H,2006ApJ...639..838P}).
These objects seem to have a nature and origin distinct from compact GCs, and
could instead be more closely related to UCDs (e.g., \citealt{2011A&A...529A.138B}),
a theme that we explore further in this paper.

In order to orient the analysis that follows, 
we need a clear observational definition for UCDs, which has varied significantly in
the literature.  
We adopt the provisional criteria that
$r_{\rm h} \sim$~10--100~pc and $M_V \la -8$ ($M_i \la -8.5$), with no upper limit
on luminosity.
These are not intended to be hard boundaries that enclose all UCDs, and only UCDs, 
but rather provide a useful outline of their primary domain. 
The unusual aspect of our definition is the extension to lower-luminosity objects
than have previously been considered as UCDs.
One of our goals in this paper is an empirical refinement of the
boundaries between UCDs and star clusters in size-luminosity space.

Given these criteria, identifying UCDs beyond the Milky Way requires 
both high-resolution imaging to determine sizes accurately
(generally from the \emph{Hubble Space Telescope} [{\it HST}]),
and distance measurements to verify sample membership 
(typically from spectroscopic redshifts).
In the past, UCD selection has often had to rely on one or the other; rarely have both critical criteria been in play to generate a large sample.

In this spirit, we revisit the population of UCDs around the giant elliptical galaxy M87, at the center of the Virgo cluster.  The new comprehensive survey of this system, as described below, has allowed us to double the sample of confirmed M87 UCDs and study their scaling relations at lower luminosities than ever before. 
This is the largest and most complete UCD sample for any galaxy studied to date, and the homogeneous
distance and environment permit an unprecedentedly accurate characterization of true UCD properties.

Our observational results are presented in Section~\ref{sec:obs}.
We analyze size trends with luminosity and galactocentric distance
in Sections~\ref{sec:lum} and \ref{sec:dist}, respectively.
We consider color, age, metallicity, and velocity trends in Sections~\ref{sec:cmd} and \ref{sec:kin}, 
and discuss UCDs in a wider context in Section~\ref{sec:disc}. 
Our summary and conclusions are in Section~\ref{sec:concl}.
Appendix~\ref{sec:uber} presents an extensive new catalog of sizes and luminosities
for nearby stellar systems, using our own data and information from the literature.

\section{Observations}\label{sec:obs}

The UCDs were drawn from a large spectroscopic and photometric survey of GCs in M87, described in detail in \citet[hereafter S+11]{Strader11}. Here we give a brief summary of the relevant points. 

Multi-color wide-field photometry was obtained with Subaru/Suprime-Cam and CFHT/Megacam,
and follow-up spectroscopy with LRIS and DEIMOS on Keck, as well as with Hectospec on the MMT.
These data yielded a new catalog of precise radial velocities for 451 compact
stellar systems,
which was combined with existing data from the literature to create a comprehensive spectroscopic compilation of 737 objects associated with M87 (extending out to a projected galactocentric
radius of $R\sim$~200 kpc, for an adopted distance of 16.5~Mpc).

Previous studies have identified 16 objects around M87 that fit our criteria for UCDs
\citep{2005ApJ...627..203H,2006AJ....131..312J,2007PhDT.........4H,2007AJ....133.1722E,2008AJ....136..461E}.
We refer to this as the sample of ``old" (prior to this work) UCDs in M87\footnote{Although
some of these objects have been called ``Virgo'' UCDs, they all have positions
($R <$~140~kpc) and velocities that are consistent with being bound to M87. 
Also, as discussed in S+11,
there is evidence that some of the older, lower-resolution spectra of objects around M87 had
``catastrophic'' velocity measurement errors of up to 1200~\kms. 
However, given the brightness of the UCDs,
it seems unlikely that this would have led to any complete misclassifications of background galaxies as Virgo objects.}.
There are another dozen bright,
spectroscopically-confirmed objects (with $M_i\sim-12$) that undoubtedly
include some bona fide UCDs,
but they do not yet have size measurements, so we exclude them from our 
confirmed sample\footnote{These objects are
H18539, H60812, H62525, VUCD8, VUCD10, S348, S784, S804, S1370, S1538, S1584, and S1617
(\citealt{1987AJ.....93..779H,1987AJ.....93...53M,1997ApJ...486..230C,2001ApJ...559..812H,2005ApJ...627..203H,2006AJ....131..312J,2008MNRAS.389.1539F,2010ApJ...724L..64P}; S+11). There is also a bright intergalactic UCD candidate, IGC1285, at a projected distance of 740~kpc from M87 \citep{2008MNRAS.389.1539F,2009MNRAS.394.1801F}.
Additional objects of potential interest are S923, which has a very high velocity relative to M87 and exhibits a peculiar asymmetric, multi-component structure;
and S7023, which may have a very low velocity and shows a core plus halo structure (see S+11).}.

For all of the objects in the spectroscopic catalog without published {\it HST}-based half-light radii \rh\ in the literature, we searched for appropriate {\it HST} archival images that were sufficiently deep for size measurements. ACS, WFPC2, and STIS images in reasonable optical bands (from essentially $B$ to $I$-equivalent) were used. Sizes were measured using \emph{ishape} \citep{1999A&AS..139..393L},
adopting a King model with a fixed concentration of 30.

It is known that some of the Virgo UCDs are not well fitted by King models, or else
have concentrations that are lower than 30, but that in many cases the inferred sizes are only modestly 
affected by the model choice \citep{2008AJ....136..461E}.  Most of the objects we
analyze do not have the very high imaging $S/N$ required to fit for more than one
free density-profile shape parameter \citep{2001PASP..113.1522C}, so we must unavoidably adopt a simple, consistent
model for all of the objects.  We also re-analyze the fainter UCDs from the 
Ha{\c s}egan samples, rather than using those authors' reported \rh\ values that were based
on more general models.  In a few cases our UCD size measurements are significantly ($\sim 30\%$) larger, underscoring the fact that \rh\ is a model dependent quantity. 

\begin{deluxetable*}{lccrrrccrl}
\tablewidth{0pt}
\tabletypesize{\footnotesize}
\tablecaption{Ultra Compact Dwarfs around M87 \label{tab:data}}
\tablehead{ID & R.A. & Decl. & $R$ & $M_V$ & $M_i$ & $(g-i)_0$ & $v$ & \rh\ & $r_{\rm h}/r_{\rm t}$ \\
& [J2000] & [J2000] & [kpc] & & & & [\kms] & [pc] }
\startdata
{\it Old confirmed:}\\
VUCD7 & 187.97040 & 12.26641 & 82.62 & $-12.6$ & $-13.25$ & 0.95 & $985\pm3$ & 100.6 & 0.087 \\
S547 & 187.73910 & 12.42903 & 14.36 & $-12.3$ & $-13.12$ & 1.15 & $714\pm2$ & 21.6 & 0.063 \\
VUCD5 & 187.79950 & 12.68364 & 88.25 & $-12.2$ & $-12.89$ & 1.06 & $1290\pm2$ & 19.2 & 0.018 \\
VUCD1 & 187.53155 & 12.60861 & 79.54 & $-12.2$ & $-12.73$ & 0.89 & $1223\pm2$ & 12.1 & 0.013 \\
VUCD2 & 187.70085 & 12.58636 & 56.24 & $-12.1$ & $-12.63$ & 0.87 & $824\pm50$ & 14.1 & 0.019 \\
VUCD4 & 187.76865 & 11.94347 & 130.12 & $-12.0$ & $-12.56$ & 0.90 & $916\pm2$ & 25.1 & 0.020 \\
VUCD6 & 187.86816 & 12.41766 & 46.26 & $-11.9$ & $-12.46$ & 0.90 & $2100\pm2$ & 18.8 & 0.031 \\
S417 & 187.75616 & 12.32351 & 24.06 & $-11.7$ & $-12.33$ & 1.00 & $1860\pm2$ & 14.7 & 0.039 \\
H55930 & 187.63929 & 12.49845 & 36.14 & $-11.7$ & $-12.28$ & 0.84 & $1296\pm4$ & 28.9 & 0.059 \\
VUCD9 & 188.06074 & 12.05149 & 139.78 & $-11.7$ & $-12.27$ & 0.91 & $1216\pm61$ & 25.4 & 0.021 \\
S928 & 187.69875 & 12.40845 & 5.38 & $-11.3$ & $-11.82$ & 0.80 & $1283\pm5$ & 36.3 & 0.30 \\
H36612 & 187.48603 & 12.32538 & 64.69 & $-11.2$ & $-11.70$ & 0.84 & $1599\pm3$ & 14.5 & 0.024 \\
S5065 & 187.70854 & 12.40248 & 3.35 & $-11.1$ & $-11.65$ & 0.86 & $1578\pm3$ & 26.1 & 0.32 \\
S999 & 187.69130 & 12.41709 & 8.53 & $-10.9$ & $-11.42$ & 0.81 & $1466\pm5$ & 33.7 & 0.23 \\
S8006 & 187.69436 & 12.40616 & 5.41 & $-10.7$ & $-11.20$ & 0.81 & $1079\pm5$ & 31.7 & 0.32 \\
S8005 & 187.69252 & 12.40641 & 5.80 & $-10.6$ & $-11.14$ & 0.81  & $1883\pm5$ & 36.9 & 0.36 
\vspace{0.1cm}
\\
{\it New confirmed:}\\
S477 & 187.74961 & 12.30030 & 28.90 & $-11.1$ & $-11.59$ & 0.74 & $1651\pm62$ & 33.5 & 0.10 \\ 
S1629 & 187.61066 & 12.34572 & 29.82 & $-10.8$ & $-11.32$ & 0.79 & $1136\pm11$ & 26.4 & 0.082\\ 
H30772 & 187.74191 & 12.26728 & 37.07 & $-10.7$ & $-11.25$ & 0.91 & $1224\pm9$ & 9.9 & 0.027 \\ 
S686 & 187.72421 & 12.47187 & 23.81 & $-10.7$ & $-11.15$ & 0.78 & $817\pm106$ & 21.2 & 0.080 \\ 
S796 & 187.71563 & 12.34815 & 12.67 & $-10.5$ & $-11.01$ & 0.79 & $1163\pm106$ & 15.3 & 0.092 \\ 
S672 & 187.72804 & 12.36065 & 10.76 & $-10.4$ & $-10.93$ & 0.78 & $735\pm106$ & 25.9 & 0.18 \\ 
S887 & 187.70389 & 12.36544 & 7.42 & $-10.2$ & $-10.76$ & 0.86 & $1811\pm106$ & 10.0 & 0.093 \\ 
S731 & 187.72452 & 12.28682 & 30.49 & $-10.2$ & $-10.72$ & 0.86 & $1020\pm9$ & 24.8 & 0.091 \\ 
H27916 & 187.71521 & 12.23610 & 44.72 & $-10.2$ & $-10.64$ & 0.76 & $1299\pm10$ & 13.7 & 0.040 \\ 
S1201 & 187.67423 & 12.39478 & 8.98 & $-10.1$ & $-10.54$ & 0.71 & $1211\pm106$ & 29.9 & 0.26 \\ 
S682 & 187.72775 & 12.33962 & 16.05 & $-9.9$ & $-10.34$ & 0.71 & $1333\pm106$ & 23.7 & 0.15 \\ 
S6004 & 187.79259 & 12.26697 & 43.28 & $-9.8$ & $-10.34$ & 0.79 & $1818\pm77$ & 40.3 & 0.13 \\ 
H30401 & 187.82795 & 12.26247 & 50.51 & $-9.7$ & $-10.16$ & 0.75 & $1323\pm46$ & 10.7 & 0.033 \\ 
S825 & 187.71263 & 12.35542 & 10.46 & $-9.6$ & $-10.10$ & 0.76 & $1142\pm106$ & 13.3 & 0.12 \\ 
S723 & 187.72399 & 12.33940 & 15.74 & $-9.6$ & $-10.03$ & 0.77 & $1398\pm106$ & 16.9 & 0.12 \\ 
H44905 & 187.73785 & 12.39440 & 9.03 & $-9.4$ & $-9.86$ & 0.77 & $1563\pm18$ & 40.0 & 0.43 \\ 
S1508 & 187.63087 & 12.42356 & 23.08 & $-9.1$ & $-9.57$ & 0.74 & $2419\pm140$ & 42.4 & 0.27 \\ 
H39168 & 188.15205 & 12.34920 & 126.07 & $-8.3$ &  $-8.76$ & 0.74 & $1349\pm13$ & 11.0 & 0.029  
\vspace{0.1cm}
\\
{\it New candidates:}\\
H46823 & 187.73054 & 12.41109 & 9.00 & $-10.4$ & $-10.91$ & 0.82 & \nodata\ & 17.0 & 0.13 \\
H46017 & 187.72083 & 12.40476 & 5.74 & $-9.4$ & $-9.85$  & 0.71 & \nodata\ & 31.5 & 0.46 \\
H42003 & 187.74030 & 12.37334 & 10.93 & $-9.2$ & $-9.75$  & 0.82 & \nodata\ & 34.3 & 0.34 \\
H46484 & 187.69745 & 12.40857 & 5.56 & $-9.2$  & $-9.66$  & 0.79 & \nodata\ & 39.1 & 0.62 \\
H41821 & 187.69752 & 12.37159 & 6.10 & $-8.9$ & $-9.58$  & 0.99 & \nodata\ & 29.4 & 0.45 \\
B1  & 187.70503 & 12.40549 & 4.15 & $-8.1$ & $-8.7$ & 0.8 & \nodata\ & 34.0 & 0.87 \\
\enddata
\tablecomments{The confirmed UCDs are grouped in ``old'' and ``new'' subsets depending on
whether they were explicitly confirmed by earlier work (via distance and size measurements)
or are new to this paper.  
Within each grouping, the objects are ordered by magnitude $M_i$.
See Section~\ref{sec:dist} for derivation of the estimated tidal radii $r_{\rm t}$.
}
\end{deluxetable*}

Uncertainties are difficult to estimate for our size measurements due to the range in instruments, depth, and filters, but random errors of $\sim$~15\% for $S/N\sim$~150 are reasonable.
This is supported by
cross-checks using different instrument set-ups and with other authors' size measurements 
(see S+11 for more details).

A combined sample of 344 spectroscopically-confirmed
M87 objects with measured sizes is reported and tabulated in S+11,
which also contains further details about their properties and the analysis methods.
With our current criteria of $r_{\rm h}\sim$~10--100~pc, we identify
34 UCDs associated with M87 and provide their details in Table~\ref{tab:data}.
Of these 34, 18 are ``new" in the sense that they have not been identified as UCDs in previous works, and are of relatively low luminosity. 15 of the 18 are completely new identifications, with the first published sizes in S+11; four also had no previous spectroscopy. Of the remaining three objects, two (S887 and H30772) have sizes from \citet{2009ApJS..180...54J}; the last (S672) has a size from \citet{2009ApJ...705..237M} but no previous redshift.  

As we will discuss later,
there are indications from colors and kinematics that the UCD population of
M87 extends to much smaller sizes than the conventional $r_{\rm h}\sim$~10~pc boundary.
For now, we designate objects in the $r_{\rm h}\sim$~5--10~pc range as ``intermediate''
objects, and smaller objects ($r_{\rm h}\la$~5~pc) as GCs\footnote{These are
not absolute limits for all objects.  E.g., some bona fide GCs could have 
evolved from originally compact sizes to be larger than 5~pc.  
However, for the relatively massive objects in M87
that will be our primary focus, the two-body relaxation timescales are too long
for the GCs to have expanded to $r_{\rm h} \sim$~10~pc.}.

An important component to any size analysis is understanding the sample selection effects.
The main, unavoidable, selection in our spectroscopic sample is by target magnitude.
Due to the composite nature of the sample drawn from several different surveys, the magnitude limits
have a complex dependency on galactocentric distance $R$ (see figure~10 of S+11).  
In brief, the spectroscopic sample with sizes available has a magnitude limit of 
$i_0 \sim 22$ ($M_i \sim -9$, $M_V \sim -8.5$) at $R \sim$~8--25 kpc, and
$i_0 \sim 21$ ($M_i \sim -10$, $M_V \sim -9.5$) at $R \sim$~25--45 kpc.
At larger distances, the sample is strongly skewed to the very bright end
($M_i \la -12$) owing to an {\it HST} program that targeted such objects \citep{2008AJ....136..461E}.

This magnitude-limit variability could potentially impact some inferences about the relations between
distance, luminosity, and size that we will examine later.
However, an important point here is that the sizes were all measured {\it after} spectroscopic
confirmation, so there should be no inherent size bias (at a given luminosity and distance).

The one caveat here is that a size criterion of FWHM $< 1.1^{\prime\prime}$ from
the Megacam imaging was used to select the spectroscopic targets and exclude
background galaxies. This imaging had seeing of $\sim$~0.7$^{\prime\prime}$, and
any UCDs in the $r_{\rm h} \ga$~50--100~pc regime may have been selected against.
Our preliminary analysis of the ground-based imaging indicates that there are very
few viable candidates of this nature, as nearly all objects above this size cut
have colors or morphologies that suggest background galaxies. However, it
should be kept it mind that this is an area of parameter space in need of more
exploration to find rare but interesting objects.

Although our sample represents a major improvement over previous M87 UCD surveys that reached
$M_i \sim -11.5$ (see references above), it is still highly incomplete. To estimate
the total numbers of UCDs, we
assume that they follow the spatial distribution of the ``blue'' GCs around M87
(because the vast majority of the UCDs have blue colors). 
We estimate $\sim$~300 bright blue ``GCs'' to be in the magnitude range
of $M_i=-10.5$ to $-12$ and in the distance range $R=$~10--200~kpc (S+11).
We have found that $\sim$~15\% of such objects are UCDs and therefore, after
allowing for UCDs with red colors and other magnitudes and distances, we expect that there
are easily another $\sim$~50 M87 UCDs awaiting discovery.

We will also be comparing our spectroscopically-confirmed sample to
a larger sample of photometrically-identified GCs from a central ACS pointing on M87 
\citep{2009ApJS..180...54J}. This allows us to increase the luminosity range surveyed, and 
to compare to a relatively unbiased sample.
This GC catalog is thought to have a very low level of contamination from 
foreground stars and background galaxies, but it
deliberately omitted any objects with \rh~$>10$ pc and $M_i \la -12.7$.
An analysis of the same dataset by \citet[figure 6]{2005ApJ...627..203H} 
shows that there are a handful of larger objects that have not been yet reported 
explicitly, and which could be UCDs.

To find these objects, we carry out aperture difference photometry on all 
pointlike objects in the ACS images (which are in bands equivalent to $g$ and $z$),
to a faint-magnitude limit of $g=23.5$ (our approximate
spectroscopic limit).  We find seven extended objects whose colors are consistent
with the known M87 UCDs and which are not visually identifiable as obvious background galaxies. One of these, H41729, was already found from a surface brightness
fluctuation analysis to be a background galaxy \citep[object 1316\_1]{2005ApJ...627..203H}.
It may be part of a cluster of galaxies behind M87 at $z=0.09$ \citep{1984ApJ...280..547H},
and provides a warning about the purity of UCD selection without distance confirmation.

The properties of the remaining six central UCD candidates are reported at the
end of Table~\ref{tab:data}.  One of them, H41821, has an unusually red color 
and appears to be fairly near to both H41729 (which is also red)
and a very large background galaxy.
We therefore assume it to be another background object and omit it from all of our plots. This leaves 5 viable UCD candidates that merit spectroscopic follow-up.

\section{Size vs.~\,luminosity}\label{sec:lum}

\begin{figure}
\vspace{0.1cm}
\epsscale{1.0}
\plotone{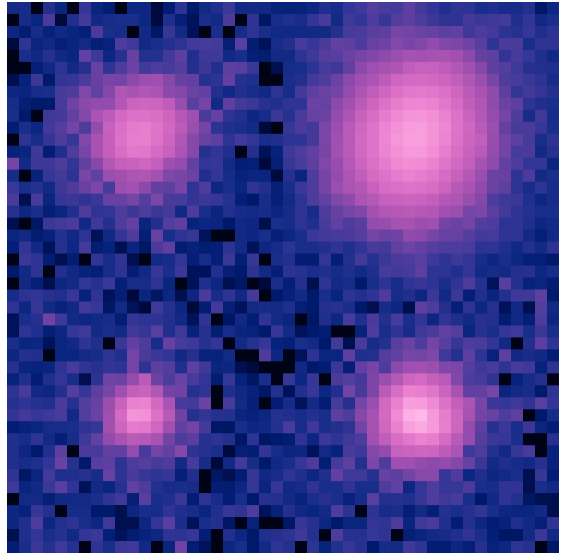}
\figcaption[ucds1]{\label{fig:thumb}
Thumbnail {\it HST} images of four representative spectroscopically-confirmed objects around M87. 
All objects are observed with the same instrument and similar filters (WFPC2 F555W and F606W),
and the images are background-subtracted with the same scaling applied.
Each thumbnail has a size of $2.3^{\prime\prime}\times2.3^{\prime\prime}$ or $183\times183$~pc
(pixel scale of $0.10^{\prime\prime}$ or 8.0~pc).
Clockwise from top-right: classic UCD (S477: $M_V=-11.1$, \rh{}~$=34$ pc), bright compact GC (S1200: $M_V=-10.8$, \rh{}~$=3$ pc), low luminosity compact GC (H45240: $M_V=-9.2$, \rh{}~$= 3$ pc), low luminosity UCD (S682: $M_V=-9.9$, \rh{}~$=24$ pc).
The final one represents a new class of object.}
\end{figure}

Most of the 18 newly identified UCDs are fainter than specified in conventional
definitions ($M_i \ga -11$, $M_V \ga -10.5$), and we have thus discovered a new area
of size-luminosity parameter space that is inhabited by UCDs\footnote{Similarly low luminosity UCD candidates were identified in the core of the Coma cluster by \citet{2010ApJ...722.1707M} but have not been spectroscopically confirmed. 
Many other spectroscopic studies of ``GCs'' around other galaxies
are bound to include some fainter UCDs (e.g., \citealt{2007MNRAS.382.1342F}), 
but these await confirmation via size measurements.
The only previously confirmed examples from this area of parameter space are
NGC~2419 from the Milky Way, and 90:12 from Fornax \citep{2005A&A...439..533R}.
Several other objects also previously appeared to have somewhat unusually large sizes relative
to UCD size vs.\,mass expectations,
including the M87 UCD S8005, as noticed by \citet{2005ApJ...627..203H}.}.
Two of the objects, H44905 and S1508, are by far the lowest surface brightness 
UCDs confirmed to date, with $\mu_V \sim 22.4$~mag~arcsec$^{-2}$ within the half-light radius.
Some example images are shown in Figure~\ref{fig:thumb}.

\begin{figure}
\epsscale{1.24}
\plotone{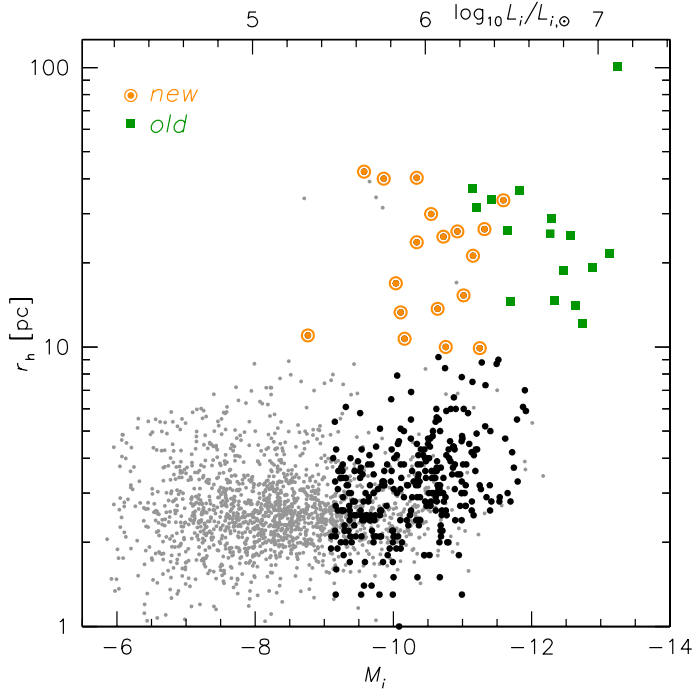}
\figcaption[M87GCsize2cw]{\label{fig:sizemag}
Half-light radius vs.~\,$i$-band absolute magnitude for compact stellar systems around M87, where all sizes are measured through {\it HST} imaging. Small gray points are from a photometric catalog of the central regions \citep{2009ApJS..180...54J}, which omitted
objects larger than \rh~$=$~10~pc, so we have supplemented it with extended objects 
brighter than our spectroscopic limit of $M_i \sim -8.5$ (see text for further details).
For this dataset,
we have estimated $i$-band magnitudes via an empirical $i-z$ vs.~\,$g-z$ calibration. 
Darker symbols are spectroscopically confirmed objects, using the entire old plus new data set: green squares are literature UCDs, while orange circles are our newly identified UCDs from S+11. The outlier at \rh~$\sim 100$ pc is the peculiar object VUCD7 which may be in a category of its own. Previous ``UCD" spectroscopy extended to only $M_i \sim -11$. The lack of a size-luminosity 
trend becomes apparent when lower luminosity UCDs are included.}
\end{figure}

We show the size and magnitude data for our full sample of M87 GCs and UCDs in 
Figure~\ref{fig:sizemag}, including for context the photometric sample of central
GC and UCD candidates as discussed in the previous section.
The addition of the new spectroscopic sample now reveals that UCDs in M87 
follow {\it no clear overall size--luminosity relation}.
In fact, the three largest UCDs in the entire sample (with \rh~$\sim$~40~pc) are
some of the faintest ones (other than the peculiar, largest object
VUCD7 which may be in a category of its own).
Curiously, there are several additional photometric UCD candidates clumped in the
same area of parameter space.

Previous claims of a strong positive size-luminosity correlation \citep{2005ApJ...627..203H,2008AJ....136..461E,2008A&A...487..921M} 
were influenced by the apparent joining of the brightest UCDs and GCs. 
Such a correlation 
was formerly linked to similar trends reported for dwarf and massive galaxy nuclei at the brighter levels \citep{2004ApJ...613..262L,2006ApJS..165...57C,2011MNRAS.414..739N} and was cited in support of the threshed nuclei origin for UCDs \citep{2008AJ....136..461E}.  
We will discuss this point further in Section~\ref{sec:cmd}.

\begin{figure}
\epsscale{1.24}
\plotone{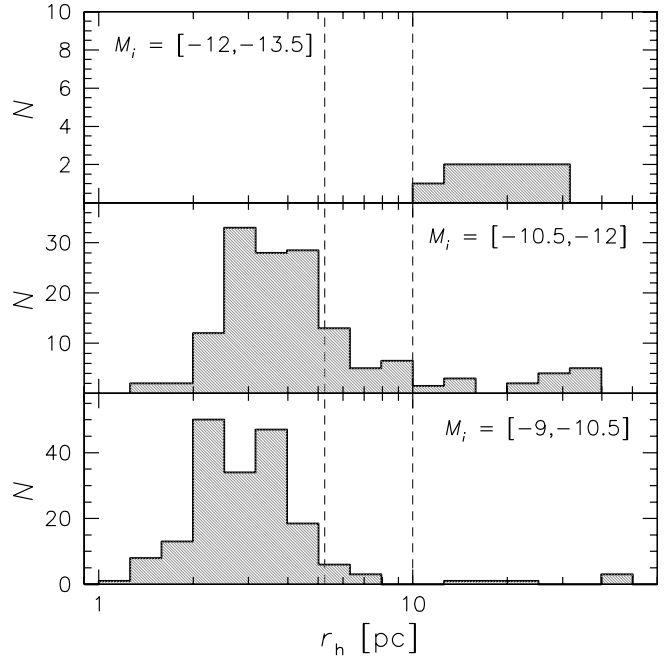}
\figcaption[M87GCsize4c]{\label{fig:hist}
Distribution of sizes for spectroscopically-confirmed compact stellar systems around M87.
The three panels show different bins in magnitude, as indicated by the labels.
The dashed vertical lines show our default boundaries between compact GCs,
intermediate-size objects, and UCDs.
The classical GCs have a narrow range of sizes peaked around $r_{\rm h} \sim$~2--4~pc, 
with a mild systematic increase with luminosity.  The UCDs sizes have a more 
uniform distribution, over a wide range ($r_{\rm h} \sim$~10--40~pc).
The proportion of compact to extended objects varies strongly with luminosity;
no GCs are found brighter than $M_i=-12$.
}
\end{figure}

The data now suggest that the UCDs and GCs occupy two separate domains in size, neither
of which has a strong luminosity dependence.
The GCs are generally smaller than $\sim$~5~pc, with a mild systematic size increase at the bright end
(also noted in other galaxies: \citealt{2006AJ....132.1593S,2007AJ....133.2764B,2009ApJ...699..254H}). 

The GC and UCD luminosities overlap over a large range of absolute magnitude, with
compact GCs of sizes $\sim$~2--3~pc found all the way up to $M_i \sim  -12$ (corresponding to $M_V \sim -11.5$ and stellar masses of $\sim 10^{7} M_{\odot}$). 
This demonstrates that identifications of UCDs by luminosity alone 
risk conflating two distinct classes of objects, except for the very brightest cases
where only extended objects are so far found.

Another way to summarize these results is as histograms of sizes in luminosity bins,
shown in Figure~\ref{fig:hist}.  With the caveat that there may be hidden correlations with
color or distance in these histograms, we note the following provisional points
(referring also to Figure~\ref{fig:sizemag}).
There is a clear ridgeline of compact GCs around $\sim$~2--4~pc, while any peak for the UCDs
is less well defined (so we continue with our $r_{\rm h} \sim$~10--100~pc UCD definition).
The peak GC size increases mildly with luminosity (by $\sim$~1~pc over a $\sim$~3~magnitude range), while the median UCD size
stays roughly constant. The fraction of UCDs varies strongly with luminosity: $\sim$~3\%, $\sim$~10\%, and $\sim$~100\%
at $M_V \sim$~$-9$, $\sim$~$-10.5$, and $\sim$~$-12$, respectively. 

\begin{figure*}
\epsscale{0.57}
\plotone{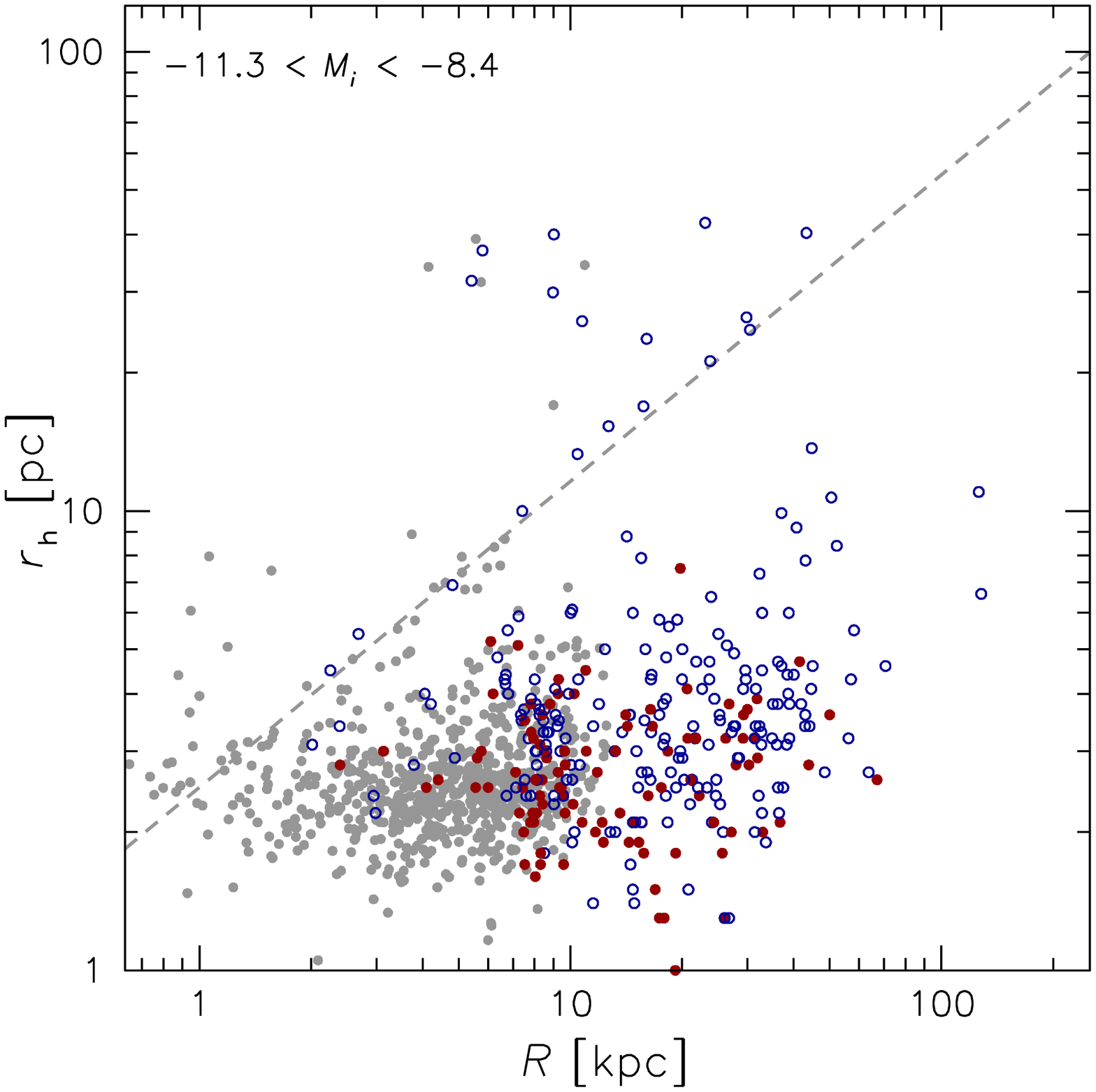}
\plotone{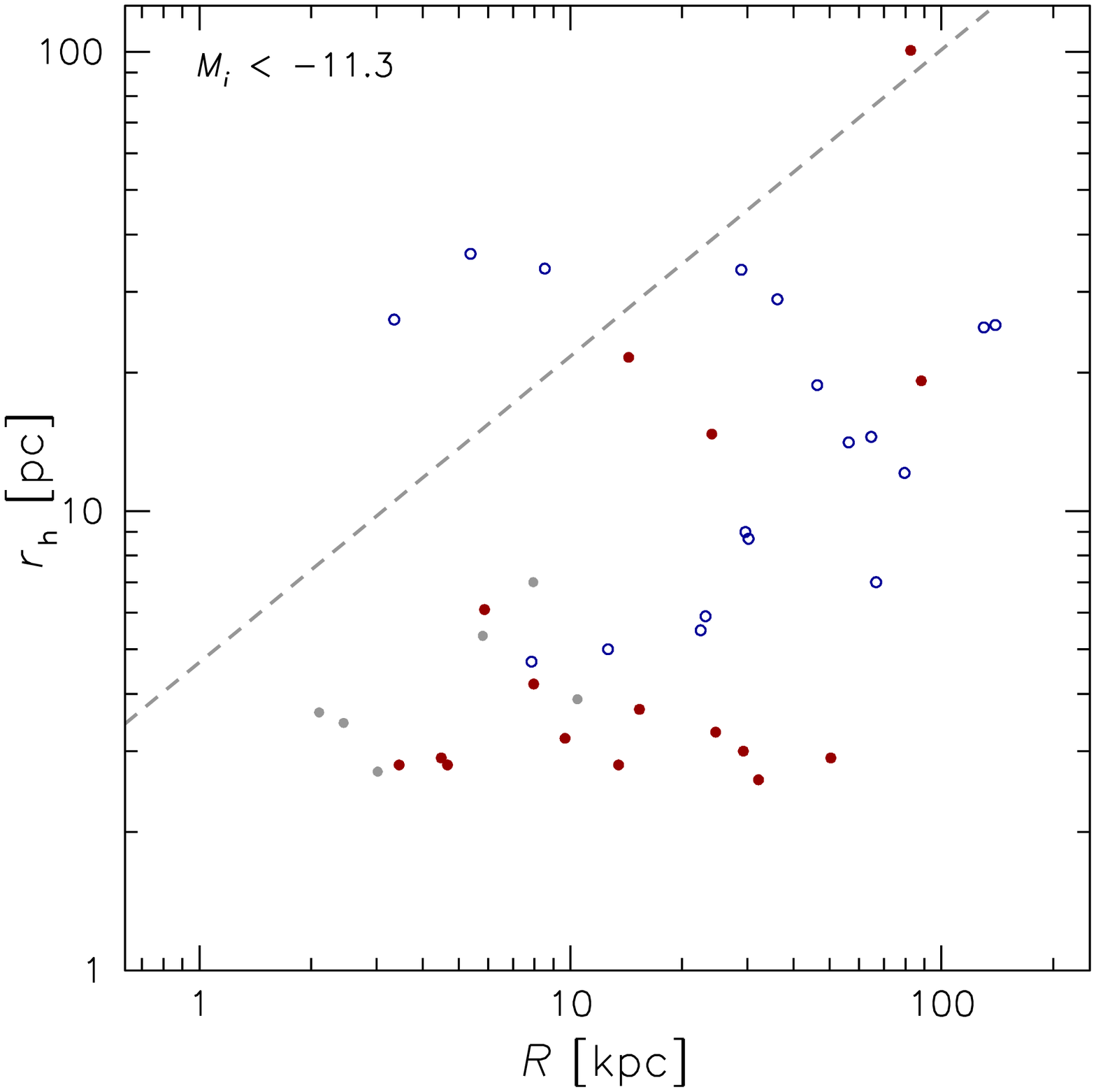}
\figcaption[M87GCsize2con4.ps]{\label{fig:dist}
As Figure~\ref{fig:sizemag}, except with
half-light radius vs.~\,projected galactocentric distance.
Open blue and filled red circles indicate assignment of the spectroscopic objects
to the metal-poor and metal-rich subpopulations, respectively.
The left and right panels show faint and bright objects, respectively,
with the luminosity ranges labeled in the panels.
The dashed line in each panel shows a tidally-limited model with a log-slope
of $\beta=2/3$, calculated using Equation~(\ref{eqn:rt})
and $M=10^6 M_\odot$ or $5\times10^6 M_\odot$ (left and right panels, respectively;
see text for details).
The GC sizes do not correlate with distance, and
the UCDs show a mild anti-correlation of size and distance.
A subset of low-luminosity GCs and UCDs may follow the tidally-limited model line.
}
\end{figure*}

\section{Size vs.~\,distance}\label{sec:dist}

Our wide-field catalog of M87 objects allows us also to explore the dependencies of size 
on galactocentric distance, $R$. 
We plot $R$ vs.~\,\rh\ in Figure~\ref{fig:dist}, where the blue and red colors correspond to the metal poor and metal rich
subpopulations well-established for GCs (e.g., Brodie \& Strader 2006),
while left and right panels show faint and bright objects. 
Note that $R$ is a projected distance, while
the more fundamental quantity is the 3-D distance.  Also, $R$ is a
measurement of the distance at a random point in each object's orbit, while the more
relevant parameter may be the pericentric distance.  Both aspects will add observational
scatter to any true correlations with distance.

Figure~\ref{fig:dist} shows a remarkable similarity to Figure~\ref{fig:sizemag} in that
compact GCs and UCDs again coexist in parallel sequences, over a wide range
of distances ($R\sim$~5--100 kpc).
Within each subpopulation, there is no clear correlation between size and distance except that
the fainter GCs seem to show a mild size increase with distance, 
from a typical \rh~$\sim$~2.5~pc inside $R\sim$~10~kpc, to $\sim$~4~pc at $\sim$~50~kpc.

Such a trend for the M87 GCs was noticed previously in the central regions
\citep{2003ApJ...593..340L,2009ApJ...703..939H,2009ApJ...705..237M}, and our data now indicate
that it extends much farther out;
beyond $R\sim$~40~kpc, no GCs more compact than $2.5$~pc are found.
There are some tantalizing similarities to size-distance trends found in other
galaxies \citep{2005A&A...442...85S,2006AJ....132.1593S,2007AJ....133.2764B,2007ApJ...668..209C,2009ApJ...699..254H,2011ApJ...736...90P,2011ApJ...738...58H}, and to the absence of compact GCs in the
Milky Way halo beyond $\sim$~25~kpc \citep{2005MNRAS.360..631M}.

Unfortunately, the reality of the distance trends in M87 is difficult to
establish because our sample selects for brighter objects in the outer regions.
A mild size-luminosity correlation could cause an apparent size-distance trend, and vice-versa.
In an attempt to reveal genuine trends, we analyze the four-dimensional space of
size, luminosity, projected distance, and metallicity, adopting a power-law model for the GCs and UCDs as follows:
\begin{equation}
r_{\rm h} \propto \, L_i^\alpha \, R^\beta \, Z^\gamma ,
\end{equation}
where we use the $(g-i)$ color as a proxy for the logarithm of metallicity $Z$.
We carry out a least-squares fit to the data for 
the GCs and UCDs separately (considering the spectroscopic sample only, and
ignoring the intermediate objects with $r_{\rm h}\sim$~5--10~pc), and estimate the uncertainties using bootstrap simulations.
We find ($\alpha=0.17\pm0.03$, $\beta=-0.02\pm0.02$, $\gamma=-0.27\pm0.06$) for the GCs, and
($\alpha=-0.03\pm0.08$, $\beta=-0.17\pm0.07$) for the UCDs (where the color diversity is
not enough to fit for $\gamma$; forcing $\gamma=0$ to the GCs for comparison does not significantly change their results for $\alpha$ and $\beta$).

These fits imply that the sizes of M87 GCs are more likely to be connected to luminosity and
metallicity than to distance\footnote{\citet{2005ApJ...634.1002J} and \citet{2010ApJ...715.1419M} 
found $\alpha$ near zero for GCs in a large sample of early-type galaxies in Virgo and Fornax,
including M87.  Our analysis applies to a relatively bright spectroscopic sample,
while for the fainter GCs we agree that $\alpha$ appears to be close to zero.}.
The GCs have
fairly constant \rh\ with distance 
at a fixed luminosity (particularly if the blue or red GCs are considered separately),
and the compact objects beyond $\sim$~40~kpc may be
missing simply because only very bright objects (which have large sizes) were sampled.
The M87 UCDs show no significant overall correlation of size with luminosity, and a mild
{\it anti-correlation} with distance\footnote{This anti-correlation may be sensitive to how the concentrations are handled in the size-fitting, but in any case a significant {\it positive} correlation appears to be ruled out.}.

We next consider the proposition that GCs or UCDs around a galaxy are tidally limited, i.e., they fill their 
tidal radius \rt.  We assume that they have fairly homologous luminosity profiles,
so that the value of $r_{\rm h}/r_{\rm t}$ is roughly the same for every GC/UCD.
The latter proposition has some support from theory \citep{2008MNRAS.389..889K,2010MNRAS.408.2353H} 
and from observations of extended clusters (ECs) in the Milky Way halo \citep{2010MNRAS.401.1832B}.

Following \citet{2010MNRAS.401.1832B}, the predicted \rt, or Jacobi radius
(which is not necessarily the same as the tidal radius from a King model fit), is:
\begin{equation}\label{eqn:rt}
r_{\rm t} = \left(\frac{4 G M R^2}{3 v_c^2}\right)^{1/3} ,
\end{equation}
where $v_{\rm c}$ is the circular velocity of the host galaxy and $M$ is the total mass
of the satellite object. We have substituted the projected distance $R$ as
a rough proxy for the 3-D distance, after multiplying by $2/\sqrt{3}$
to account for a median viewing angle of $60^\circ$.
If $v_{\rm c}$ is nearly constant with distance (as suggested by the dynamical analysis
of S+11), and $M$ scales closely with $L$, then the tidal limitation scenario predicts
the size scaling exponents $\alpha \sim 0.3$ and  $\beta \sim 0.7$.

This tidal model may explain the observed size-distance relation of Milky Way halo clusters,
with $\beta$~$\sim$~0.4--0.5
(\citealt{1991ApJ...375..594V,2000ApJ...539..618M,2003ApJ...593..340L,2005MNRAS.360..631M,2010MNRAS.401.1832B}; cf \citealt{2011MNRAS.413.2509G}).
However, the overall GCs and UCDs around M87 are inconsistent with this Milky Way finding,
with the fits suggesting little or no distance dependence.
This implies that the M87 objects'
sizes are in general {\it not} strongly influenced by tides\footnote{Projection
effects will dilute any genuine dependence of size on 3-D distance.
We have not attempted to model such effects for M87, but previous work in this galaxy's
central regions suggests that $\beta$ may be approximately halved in projection
\citep{2005ApJ...634.1002J}. 
In this case, a tidal model is still strongly excluded for the M87 GCs.}.

Examining the luminosity trends in more detail, there may be an
interesting pattern emerging for the lower-luminosity objects  
(with $M_i$ between $\sim -8.4$ and $-11.3$).
In the left panel of Figure~\ref{fig:dist},
some of the GCs and UCDs appear to coincide on a narrow diagonal track
extending from $r_{\rm h}\sim$~4~pc at $R\sim$~2.5~kpc to
$r_{\rm h}\sim$~25~pc at $R\sim$~30~kpc.
This track is mainly driven by the ACS photometric sample of GCs, which
should be complete and unbiased.
The slope of $\beta\sim0.7$ is remarkably similar to the tidal-limitation prediction.
Attempting to model this track using Equation~\ref{eqn:rt}, we adopt $v_{\rm c}=$~500~\kms,
a typical mass of $M=10^6 M_\odot$, and then adjust the tidal to half-light radius ratio
to match the data:  $r_{\rm t}/r_{\rm h}\simeq9$.

\begin{figure*}
\epsscale{0.57}
\plotone{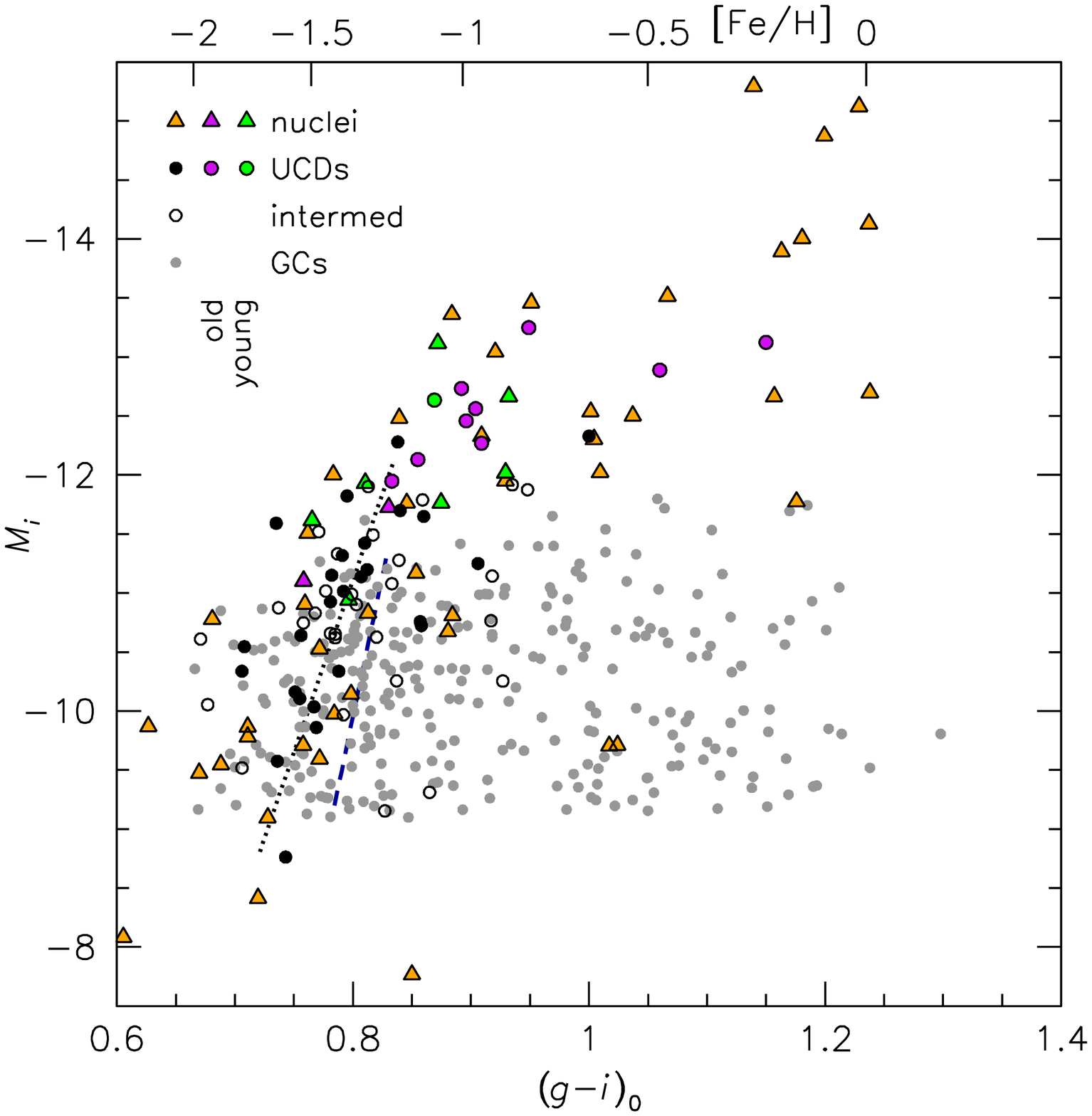}
\plotone{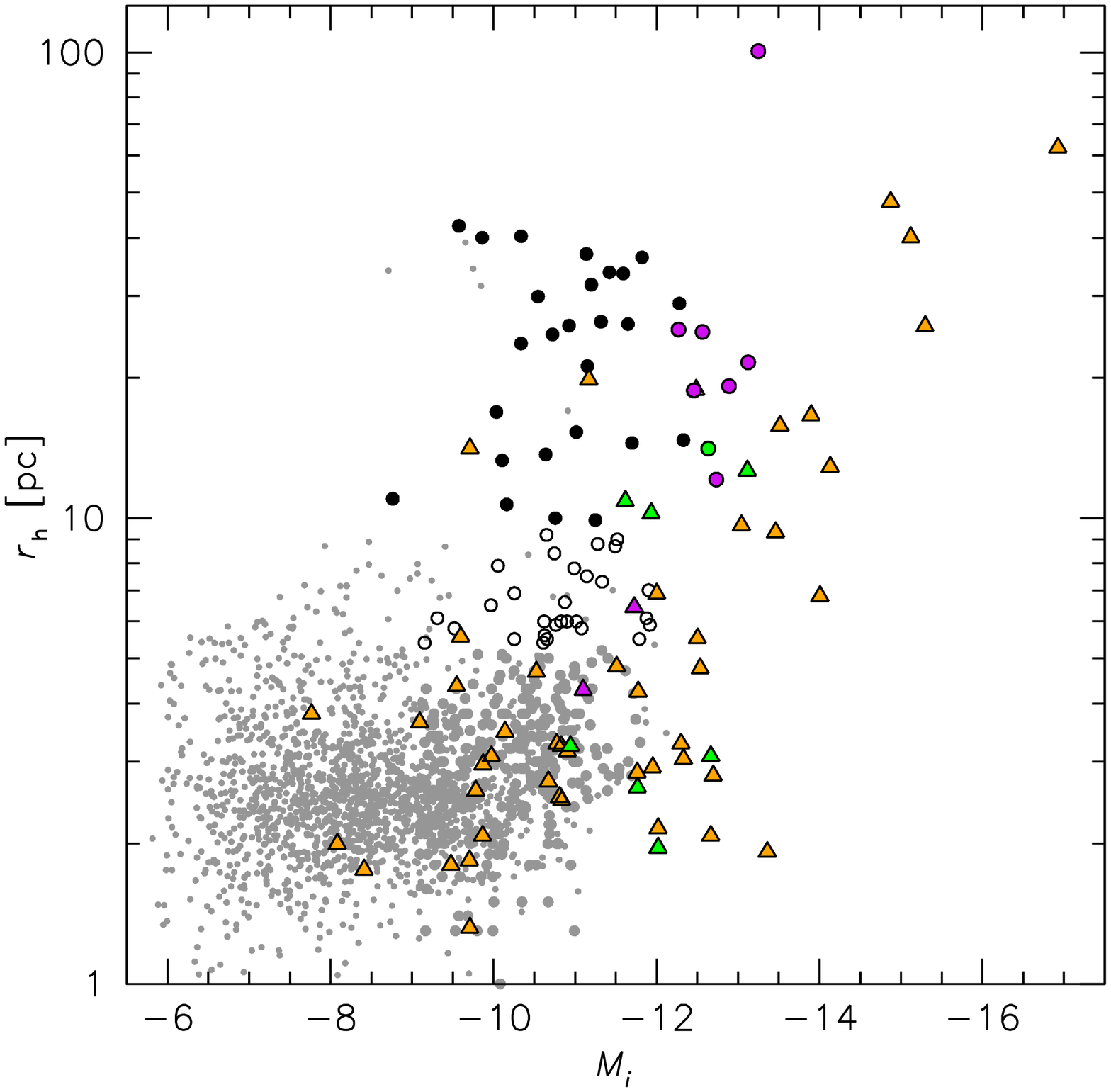}
\figcaption[M87GCcol5ag32cp2]{\label{fig:cmd}
Comparisons of dwarf elliptical nuclei in Virgo with GCs and UCDs around M87
that have been confirmed by size and redshift measurements.
Symbol types and colors are as in the legend, where
UCDs have $r_{\rm h}\ga$~10~pc,
intermediate-size objects have $r_{\rm h}\sim$~5--10~pc,
and GCs have $r_{\rm h} \la$~5~pc.
The dwarf galaxy nuclei are from \citet{2006ApJS..165...57C},
where the $gz$ photometry is converted to $gi$ using empirical relations
derived for M87 GCs in S+11.
Old and young objects are taken from the spectroscopic analyses of
\citet{2010ApJ...724L..64P,2011MNRAS.413.1764P} and are divided at 7 Gyr.
{\it Left}: Color-magnitude diagram.
The top axis shows an equivalent metallicity scale \citep{2010AJ....140.2101S}. 
The blue/red GC subpopulation boundary is at $(g-i)_0\simeq0.93$, 
or [Fe/H]~$\simeq -0.9$ (S+11).
The UCD and blue GC relations are indicated by dotted and dashed lines,
respectively, and were obtained by NMIX fitting \citep{RichGreen}.
The UCDs and dE nuclei follow
a remarkably similar and tight color-magnitude trend that is offset $\sim$~$0.03\pm0.01$
mag to the blue from the ridgeline of blue GCs.
{\it Right}: Half-light radius vs.~\,$i$-band absolute magnitude, where
central ACS photometric GC candidates (small gray points) are also included for context.
The Virgo dE nuclei are brighter for a given size than UCDs and GCs around M87.
}
\end{figure*}

The statistical robustness of this finding is unclear
(e.g., we have somewhat arbitrarily chosen the magnitude limits),
but we mention it to motivate future investigation and to
provide a tentative clue that some of the fainter UCDs may be related to
more compact GCs.  As we discuss below, the existence of such a 
subpopulation would be consistent with the notion  
of a distinct mode of diffuse star cluster formation,
producing objects whose sizes are tidally limited.

No such relationship would be inferred from consideration of the bright UCDs alone, although a 
few of these could be roughly 
consistent with the ``tidal trend'' of the faint UCDs,
after rescaling for mass (see right panel of Figure~\ref{fig:dist}).
We also provide estimated $r_{\rm h}/r_{\rm t}$ values for all of
the UCDs individually in Table~\ref{tab:data}, using Equation~(\ref{eqn:rt}) as
before, with an assumed mass-to-light ratio of $M/L_i=1.5$ in solar units.
Note that projection effects will scatter objects around a genuine tidal trend,
but there may be {\it too many} faint UCDs scattered leftwards. 
Some of these (e.g., H44905 and S8005 with $r_{\rm h}/r_{\rm t} \sim 0.4$)
would be worth more detailed follow-up to look for indications of dark matter
or ongoing disruption\footnote{Two-body relaxation should not be driving large sizes
for any of the UCDs,
as the timescales are at least $\sim$~10~Gyr for all of the objects.}.
At very small distances ($R \sim$~1--3~kpc), there are {\it no} compact objects
larger than $r_{\rm h} \sim$~8~pc,
suggesting that any extended objects in these central regions are disrupted very quickly.

Overall, there may be a population of UCDs following the scaling relations
recently suggested for ECs, 
namely $r_{\rm h} \sim 0.1 \, r_{\rm t}$
\citep[after accounting for a difference between 2-D and 3-D radii]{2010MNRAS.408.2353H}.
Ultimately, though, most of the UCDs (at all luminosities) have sizes $\sim$~4 times
smaller than this, while still being much larger than compact GCs.
This would apparently argue against a diffuse, tidally-limited mode of star cluster formation 
as the origin of these UCDs.
There is, potentially, also a problem for the stripped nuclei scenario, since these
are by definition tidally-limited objects. However,  the $r_{\rm h}/r_{\rm t}$ scaling relations discussed above
may not apply to these objects because of their initial two-component structures
(nucleus plus envelope).  

\section{Color-magnitude diagram and comparisons with dwarf elliptical nuclei}\label{sec:cmd}

In addition to direct size-related analyses of the M87 UCDs, we may survey their
other properties for commonalities with star clusters
and galaxy nuclei that could reveal shared heredity or physical influences.
These properties include color, age, and metallicity information;
and kinematics (in the next Section).

\subsection{Color-magnitude diagram}\label{sec:cmd2}

In Figure~\ref{fig:cmd} we present the color-magnitude diagram (CMD) of M87 GCs and UCDs (those with spectroscopy and measured sizes).  For magnitudes fainter than $M_i \sim -12.5$, the UCDs have a remarkably
narrow range of colors, compared to the overall GC population, or even to the blue GC subpopulation.
The faint UCD colors are slightly bluer than the blue GC peak, and follow a ``blue tilt'' of
redder colors with increasing luminosities.  Among GCs, the blue tilt has been interpreted in terms
of self-enrichment 
(e.g., \citealt{2006ApJ...636...90H,2006AJ....132.2333S,2006ApJ...653..193M,2008AJ....136.1828S,2009ApJ...695.1082B}).
At the brightest magnitudes, a few UCDs scatter to the red and it is initially
tempting to associate these with a different UCD formation channel. 

Also marked in Figure~\ref{fig:cmd} are intermediate-size objects ($r_{\rm h}$~$\sim$~5--10 pc).
Many of these lie close to the blue color sequence established by the more extended
objects, suggesting a close relationship. A revision of the UCD boundary
to sizes smaller than $r_{\rm h}\sim$~10~pc may be warranted
(we will return to this point later).

We also plot the data for nuclei from a sample of early-type Virgo galaxies \citep{2006ApJS..165...57C}.
The color trend for these nuclei tracks that of the UCDs closely, including both the
narrow blue locus and the sharp transition to red colors at bright magnitudes.
This was noticed before (e.g., \citealt{2008AJ....136..461E,2011MNRAS.414..739N}) for bright objects, and we now find that the close coincidence extends to the new low-luminosity area of parameter space for UCDs. 

A general implication is that the UCDs and nuclei have experienced very similar self enrichment processes. 
A long-standing suggestion also becomes more probable,
that UCDs have their origins as nuclei that have since been stripped by tidal forces.
The bend in the UCD color distribution could then be due to a transition from
dwarf early-type (dE) to giant early-type progenitors (e.g., \citealt{2011MNRAS.414..739N}).
Note, though, that apart 
from one extremely red object that is very large ($\sim$~100~pc), there is no particular tendency for these red objects to be larger than the average UCD. 

A narrow color spread for the blue GCs would normally imply that they are a coeval population. 
Surprisingly, the spectroscopic age estimates for a very limited subset of these M87 UCDs and dwarf nuclei (based on Lick indices; \citealt{2010ApJ...724L..64P,2011MNRAS.413.1764P}) suggest that both young and old objects in both categories conform to the same color-magnitude trends (Figure~\ref{fig:cmd}). It is a puzzling coincidence that young and old nuclei can have the same colors at a given luminosity. Accurate age determinations are notoriously difficult and the sample of objects with age estimates is small, but this intriguing result should motivate extending age studies to larger samples of both UCDs and dwarf nuclei.  
Nonetheless, the color offset between the UCDs and the blue GCs 
underscores a distinction between these two populations and again argues against a UCD origin
from star clusters, or mergers of star clusters, that
were analogous to the GCs that survive today.

In the right panel of Figure~\ref{fig:cmd} we compare the size-luminosity parameter space
for the same objects that were plotted in the left panel.
Overall, the dE nuclei are systematically brighter at a given size than the UCDs. If the brighter nuclei are found to be young, this luminosity offset might be simply explained as a result of fading (by $\sim$~2 mags). This would imply that some of the compact ``GCs'' are really ``UCDs'' in the sense that they are stripped nuclei
(e.g., \citealt{1993ASPC...48..608F}).

\begin{figure*}
\epsscale{0.57}
\plotone{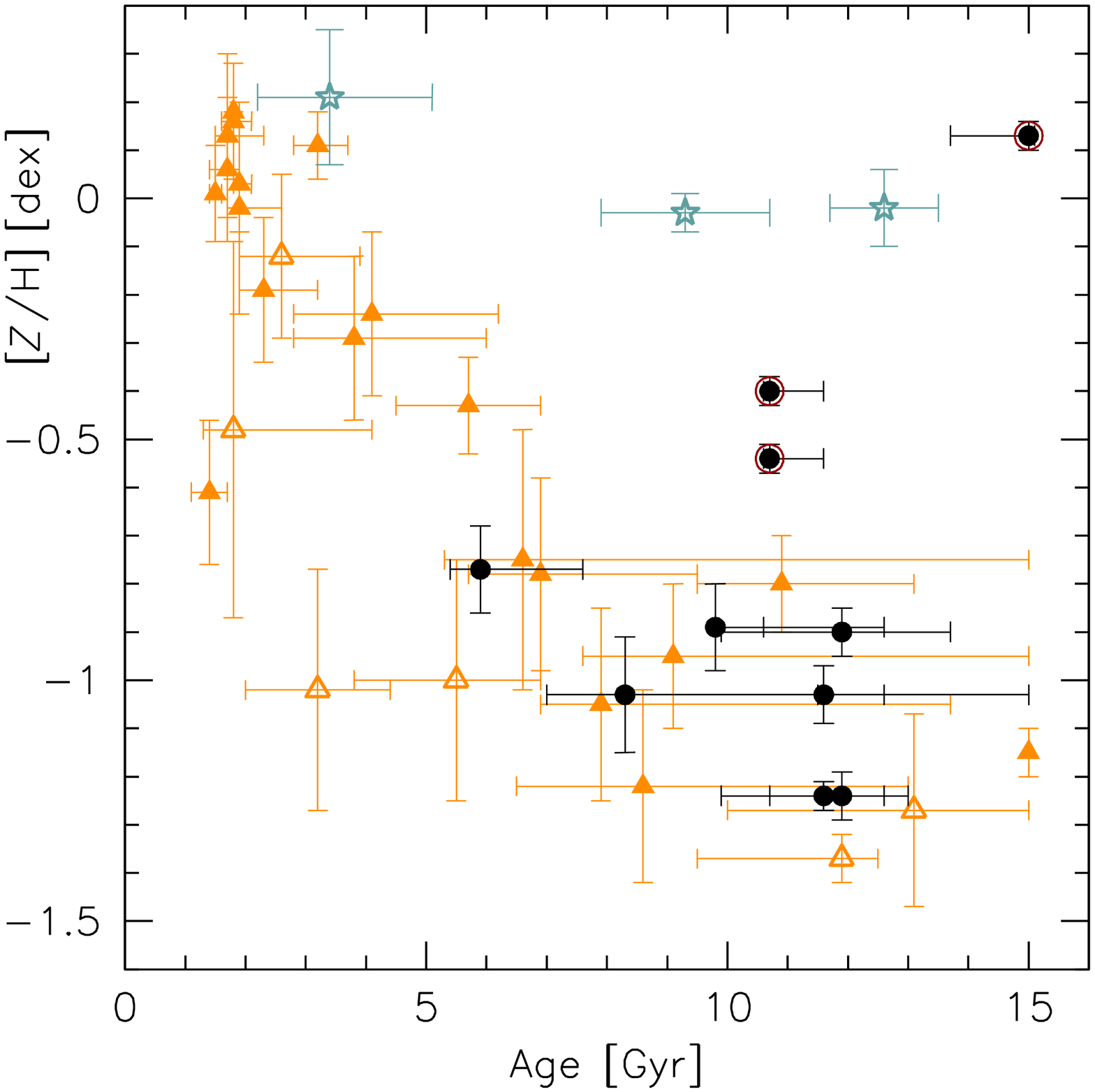}
\plotone{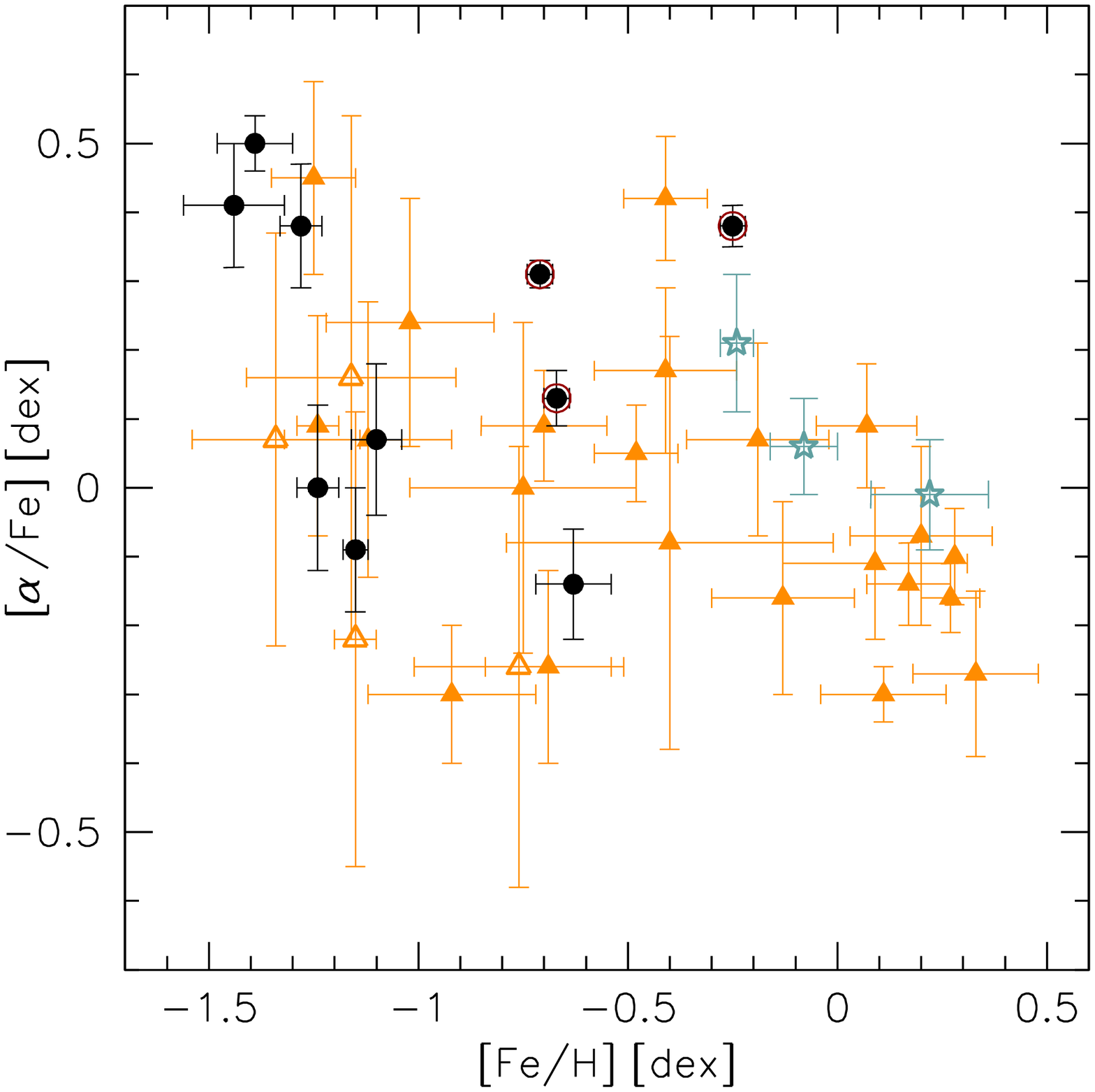}
\figcaption[M87GCcol5akl1.ps]{\label{fig:amr}
Spectroscopic populations analysis results for M87 UCDs and Virgo dE nuclei,
from \citet{2010ApJ...724L..64P,2011MNRAS.413.1764P}.
UCDs are black circles, and nuclei are orange triangles.
The three brightest UCDs (see left panel of Figure~\ref{fig:cmd}) are outlined
with red circles, while the nuclei fainter than $M_r = -11$ are shown as open triangles
(no faint UCDs were observed by Paudel et al.).
Blue star symbols mark three other UCDs with measured ages and metallicities 
\citep{2008MNRAS.385L..83C,2009MNRAS.394L..97H,2011MNRAS.414..739N}.
{\it Left:} Age-metallicity relation.  The error bars give only a rough idea of
the uncertainties, which are strongly correlated between age and metallicity owing to
degeneracies in the modeling.
{\it Right:} $\alpha$-element enhancement vs.~\,metallicity.  The [Fe/H] uncertainties
are assumed to be the same as the [Z/H] uncertainties.
Most of the M87 UCDs follow similar trends in age and metallicity to the dE nuclei.
Three of the M87 UCDs, as well as three other UCDs from the literature, are
offset in these diagrams and may originate from the nuclei of giant galaxies.
}
\end{figure*}

An alternative way to view the data is that the UCDs are larger than nuclei at a given
luminosity, which could then be interpreted as post-stripping expansion
(e.g., \citealt{2008AJ....136..461E}).
It may be that the nuclei both expand and fade after stripping, although one might then
wonder why more of the bright UCDs have not been found with sizes in the 
$r_{\rm h}\sim$~30--100~pc range.

Again, a larger number of accurate age estimates would clarify the situation.
To this end, we have carried out a preliminary analysis using  environmental density
as a proxy for age, since there appears to be a strong correlation between density
and nucleus age \citep{2010ApJ...724L..64P,2011MNRAS.413.1764P}.
We find indications that the ``older'' nuclei may have faded by only $\sim$~1~mag,
leaving expansion as a requirement to match the UCD sizes.

In more detail, the brighter nuclei ($M_i \la -13$)
show a strong size-luminosity correlation, which flattens out at lower luminosities.
Whether or not the UCDs and GCs follow the same type of trend (with a luminosity offset)
is not clear, particularly with the luminosity-dependent selection effects in the 
current sample.  Even with ideal data, the interpretation would be complicated by the 
current theoretical uncertainty about the size evolution of nuclei after stripping.
Therefore it is difficult at this point to draw firm conclusions about the UCD
origins from size-luminosity trends.

Figure~\ref{fig:cmd} also provides a useful point of comparison with the joint
size-luminosity and color-magnitude analyses of several GC/UCD systems
carried out by \citet{2011MNRAS.414..739N}.
They claimed a transition luminosity or ``scaling onset mass'' in both parameter spaces,
above which strong blue-tilt and size-luminosity relations set in
(similar transitions have been found in metallicities, velocity dispersions,
and mass-to-light ratios; e.g., \citealt{2005ApJ...627..203H,2006AJ....131.2442M,2008A&A...487..921M,2007A&A...469..147R}).
This luminosity of $M_V \sim -10$ corresponds to $M_i \sim -10.5$ in our plots, where
we see no evidence of such a transition in M87.
It could be that this luminosity is where the proportion of UCDs and GCs varies rapidly
(Figure~\ref{fig:hist}), with neither population on its own having strong trends in
size-luminosity, etc.
It is also an open question as to whether or not a significant population of UCDs exists
at magnitudes fainter than $M_i \sim -10$.

\subsection{Age and metallicity relations}\label{sec:amr}

Moving on to another area of parameter space,
Figure~\ref{fig:amr} shows the spectroscopy-based stellar populations analysis
of M87 UCDs and Virgo dE nuclei from \citet{2010ApJ...724L..64P,2011MNRAS.413.1764P}.
Before comparing these objects, it should be kept in mind that these UCDs are all found
in the high-density surroundings of M87, while the nuclei are drawn from a broad range
of environmental densities within Virgo.
Given the age-density correlation already mentioned for nuclei,
the fairest point of comparison is between the UCDs and the {\it older} nuclei.

After taking this aspect into account, and keeping in mind that the luminosity-weighting
in such spectroscopic analyses makes comparisons difficult to interpret, we notice a broad
correspondence between the UCDs and dE nuclei in their age-metallicity relations (AMRs)
and $\alpha$-element abundance distributions (as a function of mean metallicity).
The AMR for the UCDs and dE nuclei somehow conspires to result in a narrow color-magnitude
track for both classes of object (Figure~\ref{fig:cmd}, left panel).

These comparisons reinforce the suggestion from the CMD that most of the UCDs originate
from dE nuclei.  
Notice, though, the three separately tagged UCDs in Figure~\ref{fig:amr}. 
These correspond to the bright, red UCDs that we linked earlier to the stripped remnants of more massive galaxies; they are again quite distinct from the general run of dE nuclei and UCDs.

Another intriguing result from Figure~\ref{fig:amr} is that the AMRs of the faint and bright dE nuclei may be systematically different in the sense that the faint nuclei have lower metallicity at a given age (equivalent to a mass-metallicity correlation at each age) 
and lower [$\alpha$/Fe] at a given metallicity. 
The trends for the UCDs are not clear from the existing data. Spectroscopic analyses
are needed in particular for the new class of low-luminosity object, as well as for
ordinary compact GCs around M87.
 
\begin{figure*}
\epsscale{0.57}
\plotone{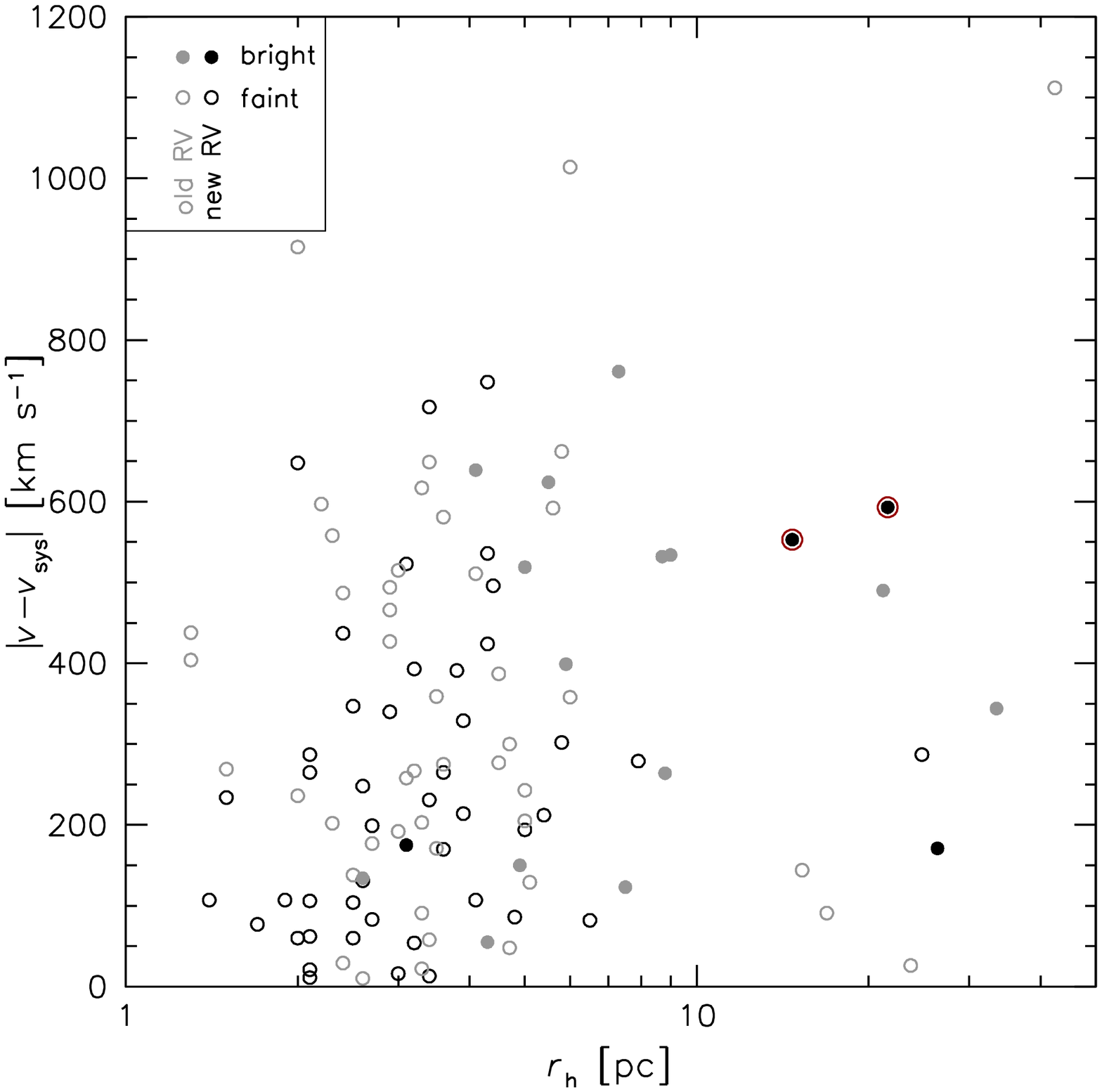}
\plotone{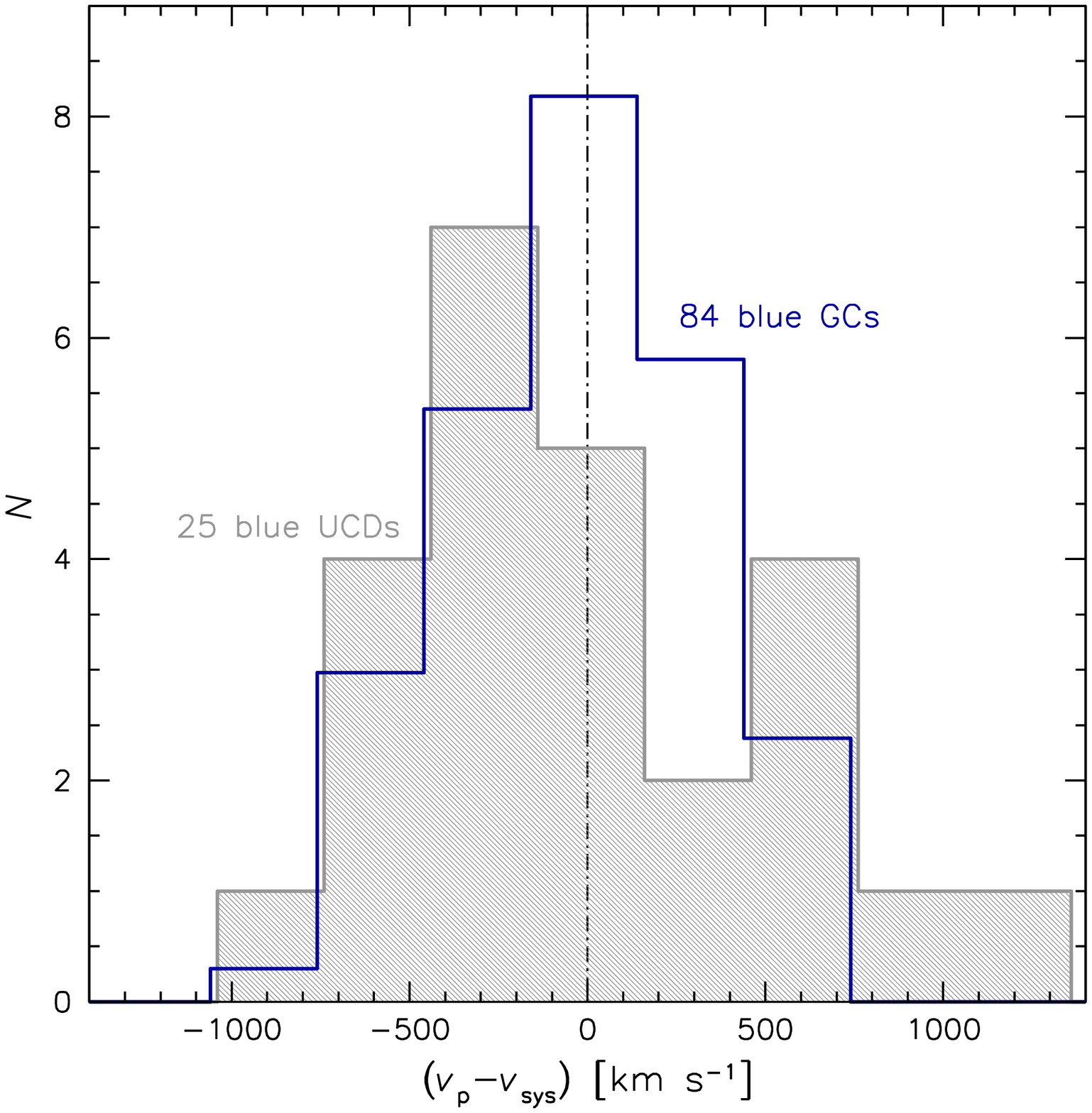}
\figcaption[M87GCsize6bvel6bu]{\label{fig:kin}
Line-of-sight velocities of blue GCs and UCDs around M87, relative to the
systemic velocity.  Objects from a distance range of $R\sim$~12--35~kpc are included.  
{\it Left}:
Trend of half-light radius vs.~\,velocity.  
The symbol types are summarized by the legend:
filled and open circles are bright and faint objects, respectively 
(with $M_i = -11$ as the boundary);
black symbols (both open and filled) show the objects with new velocity measurements (from S+11),
which have a median uncertainty of 28~\kms; gray symbols show old velocities, which have
a median uncertainty of 106~\kms, and the potential for a few ``catastrophic''
errors (see S+11) that makes the extreme velocities at $\sim$~1000~\kms\ somewhat suspect.
Two bright, red UCDs (see Figure~\ref{fig:cmd}, left panel) 
are outlined with red circles; the other two from our sample are outside of the radial range.
{\it Right}: Overall velocity distributions in the same region
for UCDs and blue GCs (gray shaded and blue open histograms, respectively), where
the size division is now at $r_{\rm h}\sim$~5~pc.
The GC histogram has been renormalized to the same area as the UCD one for the sake of comparison.
The UCDs and intermediate-size objects are less concentrated around the systemic velocity
than the GCs, resulting in a higher velocity dispersion.  
The difference may be stronger for the brighter objects.
}
\end{figure*}

Further discussion of the stellar populations implications in a wider context
will be provided in Section~\ref{sec:disc}.

\section{Kinematics}\label{sec:kin}

Understanding the origins of UCDs will ultimately require information in addition to
size, luminosity, age, and metallicity trends.
Two additional discriminators are their spatial and velocity distributions
(which are both projections of an underlying {\it orbital} distribution).
UCDs that are tidally stripped nuclei may be expected to reside on preferentially
radial orbits that result in a centrally-concentrated number density distribution,
a projected velocity dispersion profile that declines strongly with distance,
and a peaky, broad-winged shape to their line-of-sight velocity distribution
(see \citealt{1994ApJ...431..634B,2003MNRAS.344..399B,2007MNRAS.380.1177B,2008MNRAS.385.2136G,2008MNRAS.389..102T}).
Alternatively, UCDs that formed as extended star clusters might show an increasing
dispersion profile and a flat-topped velocity distribution\footnote{These predictions
are discussed in more detail in S+11.
Note that on the mass and distance scales involved here, dynamical friction should not
be a significant effect, and any kinematical peculiarities should instead be
related to orbital dependencies of the formation process.}.

The density distribution of the M87 UCDs will require further analysis that
carefully considers selection effects, but their kinematics have been analyzed in 
detail in S+11
and add to the indications from the color-magnitude diagram that the UCDs are distinct from the general GC population, and not simply a tail of the GCs to large sizes\footnote{A similar conclusion was reached by \citet{2009AJ....137..498G}, based on differences (although in the opposite sense to those found here) in mean velocity and velocity dispersion between bright and faint GCs and UCDs around NGC~1399 (no size information was used).}.

Briefly, over the distance range of $R \sim$~10--35~kpc (where the data are available
for both UCDs and compact GCs),
the UCDs and intermediate-size objects
show a broader, flatter distribution of recession velocities than the GCs
(considering only the blue GC subpopulation for a fair comparison). 

To illustrate this point further, we plot velocity vs.~\,size in Figure~\ref{fig:kin},
where the UCDs, intermediate-size objects, blue GCs are shown together.
The velocity distribution of compact objects appears to have the expected Gaussian distribution,
but the larger objects show a tendency to avoid the systemic velocity.
This behavior seems to set in for $r_{\rm h} \ga$~5~pc, supporting our suggestion from
color considerations (Section~\ref{sec:cmd2})
that many of the intermediate-size objects should be identified as small UCDs.  

With 5~pc as the GC-UCD boundary, the velocity dispersions of the GCs and the UCDs are
$340\pm30$~\kms\ and $500\pm90$~\kms, respectively.
A Kolmogorov-Smirnov test finds that the 
velocity distributions are different at the 70\% confidence level.

S+11 discussed the M87 UCD velocities in more detail, including the trends with distance
and luminosity.
The UCD velocity dispersion profile remains constant, and the
shape of the velocity distribution changes in a complicated way, neither of
which is uniquely and straightforwardly explained by either of
the formation scenarios under consideration.
It is possible that the blue UCDs comprise a mix of two different populations,
with objects of dE nuclei and star-cluster origins becoming more dominant
at  the bright and faint ends of the luminosity range, respectively.
Further theoretical work and better statistics are needed to draw firmer
conclusions about UCD origins from kinematics.

\section{Discussion}\label{sec:disc}

Up to this point, we have focused on the M87 GC/UCD system as
a high-quality, well-characterized, homogeneous dataset from a single environment.
Now we seek to understand UCDs in a broader context, 
using literature data and results from other systems.
We start by examining basic trends in size and luminosity, and then attempt to
survey a broad range of their properties in order to converge on
an integrated view of their formational histories.

To orient the discussion, we consider a basic though non-exhaustive set of four
formation scenarios for UCDs.  The first is that they are ``giant GCs'', an extension
of the normal GC population to very high masses, which naturally lead to large
sizes owing to mass-dependencies of formation or internal evolution
(e.g., \citealt{2009ApJ...691..946M,2010MNRAS.408L..16G}).
The second is that they are produced by normal star clusters that have collided
(e.g., \citealt{2002MNRAS.330..642F}). We refer to these as merged GCs.
The third is that they pertain to an independent mode of diffuse star cluster formation 
that includes the lower luminosity ECs (e.g., \citealt{2011A&A...529A.138B}).
The fourth is that they are stripped galactic nuclei
(e.g., \citealt{2001ApJ...552L.105B,2008MNRAS.385.2136G}).

Two ``smoking guns" provide direct evidence that UCDs can form in at least two distinct
ways:
W3 is likely a merged GC \citep{2004A&A...416..467M,2005MNRAS.359..223F},
and NGC 4546 UD1 is likely a stripped nucleus \citep{2011MNRAS.414..739N}.
Below, as we review the sundry properties of UCDs, we will comment at each stage
on the compatibility of the data with these different formation scenarios, and
then try to tie together the various lines of evidence into an integrated picture
of UCD origins.

\subsection{Size and Luminosity Comparisons}

We assemble from the literature a compilation of the sizes and luminosities of hot stellar systems, from the largest galaxies to the smallest GCs.
We restrict the sample to objects with distances confirmed either by spectroscopy, surface brightness fluctuations, or resolved stellar populations.
In order to not complicate the analysis with large stellar mass-to-light ratio variations,
we also include only {\it old} objects (ages $\ga$~5~Gyr) where possible,
e.g., excluding some young, extended clusters that are known in the LMC and beyond.
The results are plotted in Figure~\ref{fig:uber}, with 
the details, references, and full data table provided in Appendix~\ref{sec:uber}.

\begin{figure*}
\epsscale{1.0}
\plotone{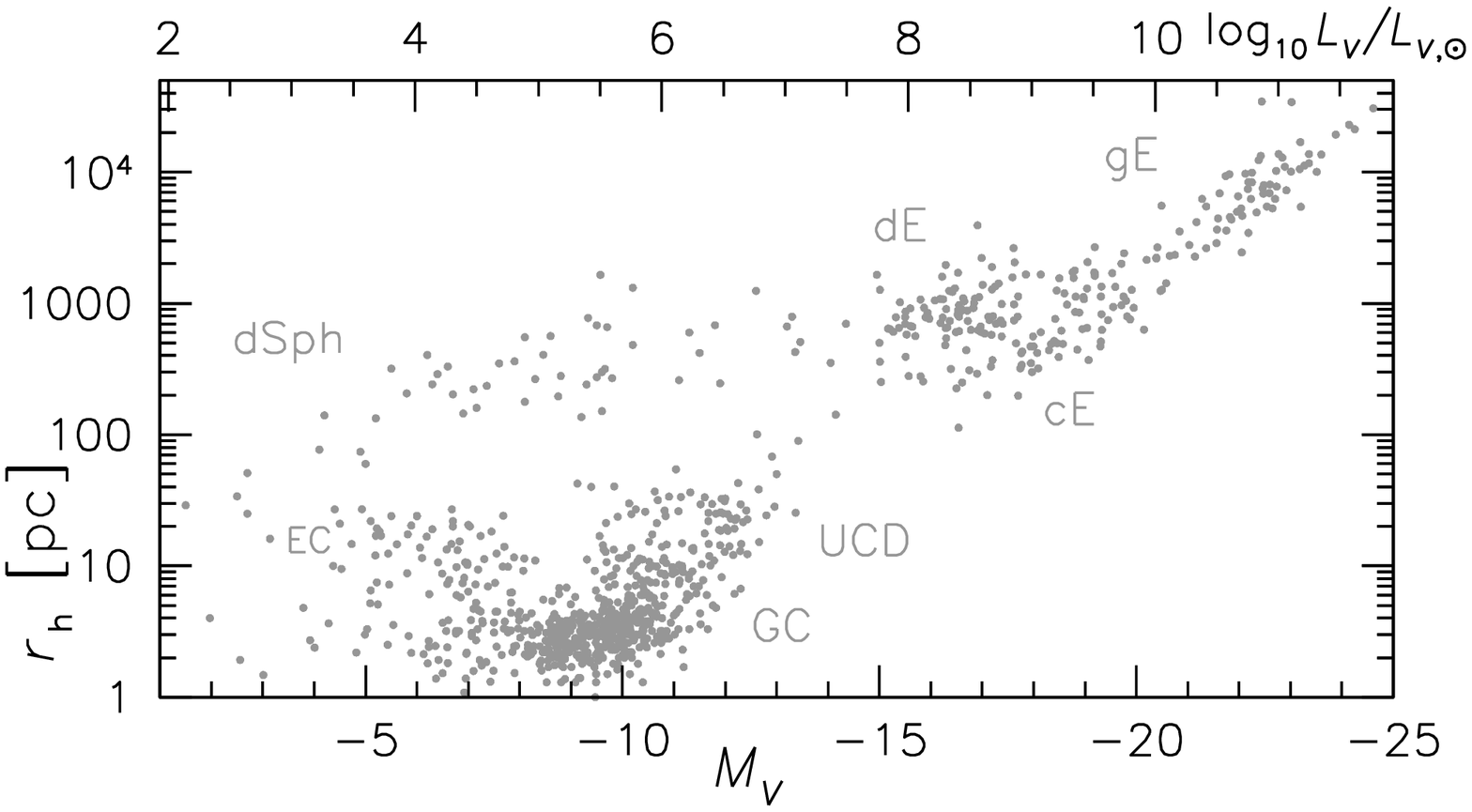}
\plotone{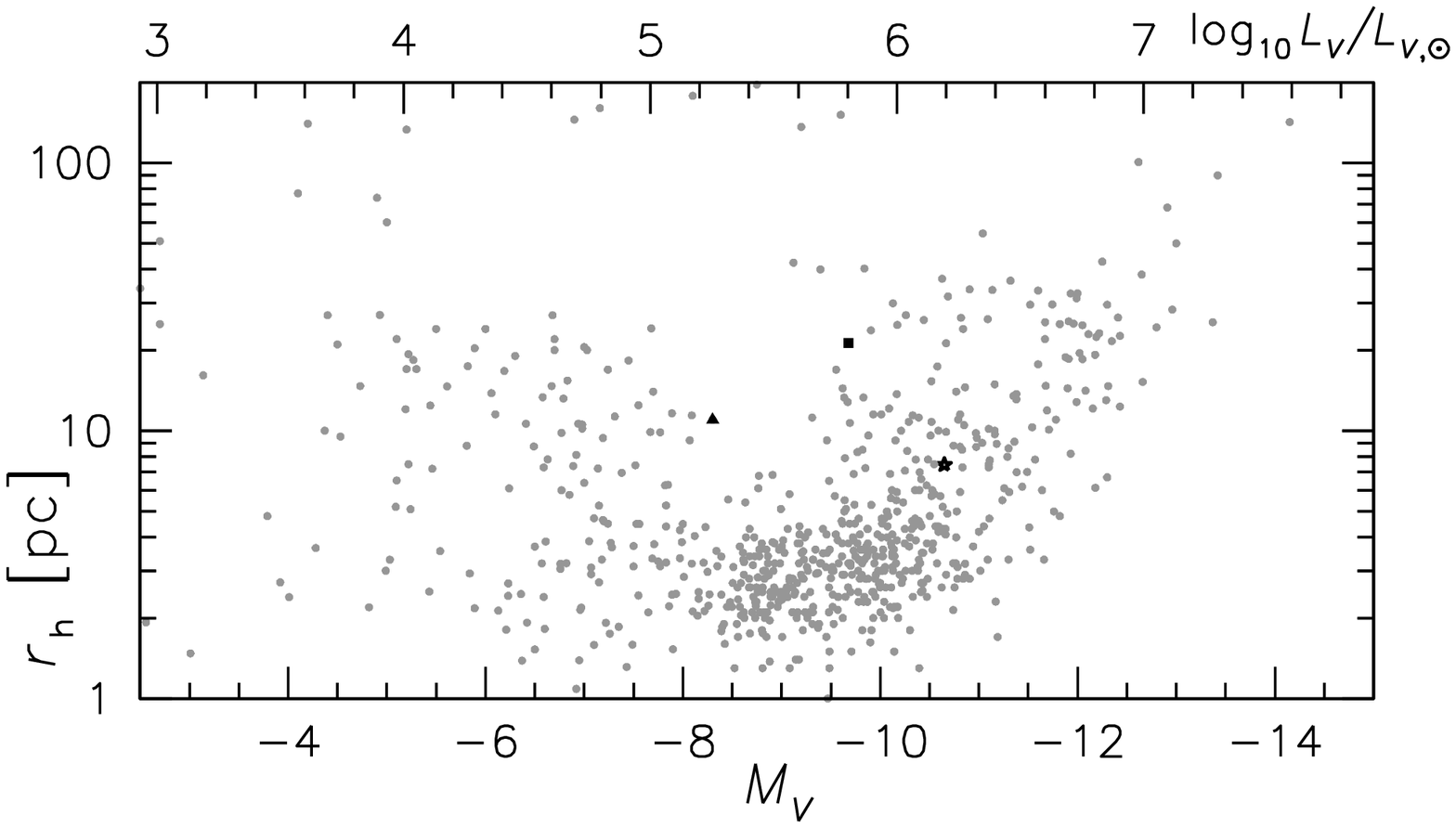}
\figcaption[M87GCsize2cqr]{\label{fig:uber}
Relations between luminosity and half-light radius for 970 stellar systems including GCs, UCDs, ECs, dwarf spheroidals (dSph), dEs, ultra-faint dSphs, giant ellipticals (gE), and compact ellipticals (cE). 
The catalog is presented in Appendix~\ref{sec:uber}, and as described there
has undergone careful quality control to prevent false outliers.
The bottom panel is a zoom-in of the top.
The cEs are likely to be stripped/threshed remnants of larger galaxies (e.g., \citealt{1973ApJ...179..423F,2011MNRAS.414.3557H}); the most famous example is M32, which is the most compact cE in the plot. 
Two well-known Milky Way objects, NGC~2419 and $\omega$~Cen, are marked for context
(black square and star symbol, respectively),
as well as the new M87 object H39168 (black triangle).
A diagonal gap is seen between the galaxies and star clusters, except where it may be
bridged by a few objects between the UCDs and cEs, which suggests that some of
the UCDs may be stripped-down galaxies.
The UCDs do not show a clear size-luminosity correlation, once one takes into account
the strong selection effects for this sample,
with only the more luminous ones ($M_V \la -11.5$) usually observed;
in addition, there may be a size-dependent upper-envelope on luminosities that
corresponds to a maximum surface density for stellar systems
\citep{2010MNRAS.401L..19H,2011MNRAS.414.3699M}.
A luminosity gap between ECs and UCDs persists, but may be produced by observational
selection effects.
}
\end{figure*}

There are various interesting features in the size-luminosity parameter space.
The most compact objects include the classical population of GCs with
$r_{\rm h}\sim$~2--4~pc, the UCDs which extend up to $r_{\rm h}\sim$~50~pc,
and the fainter ECs up to $r_{\rm h}\sim$~25~pc, which comprise not only extended clusters 
\citep{2005MNRAS.360.1007H,2008MNRAS.385.1989H,2009ApJ...698L..77H,2011MNRAS.414..770H,2006ApJ...653L.105M,2008AJ....135.1482S,2010MNRAS.404.1157M,2011ApJ...730..112C,2011ApJ...738...58H}, but also
faint fuzzies \citep{2000AJ....120.2938L,2002AJ....124.1410B,2005A&A...442...85S,2006ApJ...638L..79H,2008AJ....135.1567H,2007A&A...467.1003C,2007A&A...469..925S},
``diffuse star clusters" \citep{2006ApJ...639..838P}, and the Palomar clusters in the Milky Way halo.

There is then an apparent gap between galaxies and star clusters, which is now seen to be a diagonal region rather than a simple size gap (cf the ``Shapley line'' of \citealt{2008MNRAS.390L..51V}; and also \citealt{2007ApJ...663..948G,2008MNRAS.389.1924F,2009ARA&A..47..371T,2011MNRAS.414.3699M}). 
This gap corresponds roughly to a line of constant surface brightness,
$r_{\rm h} \propto L^{-1/2}$, so there may be a selection effect at work here,
with deeper imaging and spectroscopic surveys needed.
There are a few, possibly rare, ``bridging'' objects between the UCDs and the cEs,
which as we discuss below, may imply that star clusters and galaxies are
not completely distinct populations.

Considering the UCDs and
noting the typical strong selection biases against objects fainter than $M_V \sim -11.5$,
we see that the data suggest a nearly flat size trend that parallels the compact GCs.
The previous paradigm of a strong size-luminosity trend for UCDs is trumped by
the new discoveries of low-luminosity UCDs, mostly from around M87 but also from
a few other galaxies.  This includes the Milky Way where the halo cluster
NGC~2419 was long thought to be a unique object, 
while it can now be seen as a harbinger of the new class of UCDs
(see black square in the lower panel Figure~\ref{fig:uber}).

We thus see that standard identifications of UCDs by luminosity alone are inadvisable,
as they coexist with compact GCs over a factor of $\sim$~15 range in luminosity.
The best that can be done in the absence of direct size information is to estimate
the probability of a given object being a UCD or a GC (cf Figure~\ref{fig:hist}),
with objects more luminous than $M_V \sim -12.5$ fairly safely designated as UCDs.

Interestingly, both giant GCs 
{\it and} stripped nuclei are expected to follow strong size-luminosity (\rh-$L$) relations and
yet we find no such relation for the M87 UCDs. Upon closer scrutiny over the full range of luminosities,
the \rh-$L$ relation for nuclei is not particularly strong
(see Section~\ref{sec:cmd2}). The lack of an \rh-$L$ relation for UCDs is, however, a stronger discriminant against
the giant GC formation scenario. Here theory predicts that an \rh-$L$ relation will be
imprinted by the physics of massive cluster formation.
For merged GCs, a range of sizes are possible at a given luminosity
\citep{2011A&A...529A.138B}.

Another observation from Figure~\ref{fig:uber}
is a hint of a gap in the UCD size distribution at
$r_{\rm h}\sim$~15--20~pc (see also middle panel of Figure~\ref{fig:hist}), 
which was previously noticed for the UCDs 
near the center of a galaxy cluster by \citet{2008AJ....136.2295B}.
Such a feature might imply different modes of UCD formation, and will merit
further examination with larger samples.

There also appears to be a {\it magnitude} gap between UCDs and ECs
from $M_V \sim$~$-8$ to $\sim$~$-9$
(also called the ``avoidance zone'' by \citealt{2011ApJ...738...58H}).
Since NGC~2419 has often been proposed as an accreted nucleus 
(e.g., \citealt{2005MNRAS.360..631M,2010ApJ...725..288C}),
and ECs are normally seen as bona fide star clusters,
it is natural to interpret this as a real gap between nuclei and clusters.
In this case, H39168 from our sample would be the most luminous EC ever confirmed
($M_V = -8.3$), and S1508 would be the least luminous UCD ($M_V=-9.1$).

It is also possible that there is a continuum of extended objects,
from ECs to UCDs (where NGC~2419 has alternatively been viewed as a star cluster), 
and that the gap is an area of parameter space 
that has simply been inadequately surveyed
around nearby galaxies (cf \citealt{2006BASI...34...83R}).
While ECs in the Milky Way halo have often been identified as ordinary GCs in a late
expansion stage of their evolution (e.g., \citealt{2010MNRAS.408L..16G}),
there is also an emerging picture of two universal modes of star cluster formation,
compact and extended 
\citep{2008ApJ...672.1006E,2009A&A...498L..37P,2009AJ....137.4361D,2010MNRAS.401.1832B}. 
Given such a scenario, it is plausible that at least some of the ECs and UCDs share
a similar origin (e.g., \citealt{2011A&A...529A.138B}).

The mix of UCD origins may change with luminosity, as we have found hints that some 
of the fainter objects are star clusters (see also \citealt{2005ApJ...627..203H})
that may be linked to the even fainter class of ECs.
Many of these objects may have formed around dwarf galaxies that were later accreted
and stripped by larger galaxies.

Note, though, that ECs may themselves not be a homogeneous class.  They include, for example,
faint fuzzies that are very metal-rich and $\alpha$-enhanced, and could be associated
with mergers of either galaxies or young star clusters
\citep{2002AJ....123.1488L,2005ApJ...628..231B,2009ApJ...702.1268B}.
On the other hand, the M87 objects that may be tidally-limited (Section~\ref{sec:dist})
are generally metal-poor (blue), as is the local faint-UCD analog, NGC~2419.

Focused study on objects bordering the current EC-UCD luminosity gap may reveal 
whether or not they are related.  E.g., is H39168 (black triangle in Figure 8)
around M87 the brightest of the fuzzies
or the ``wimpiest" of the UCDs? Other fairly nearby candidate bright-fuzzies/wimpy-UCDs, besides those around M87, include
NGC~2419, NGC~361 in the SMC, 90:12 around NGC~1399
\citep{2005A&A...439..533R}, and C1 around NGC~5128 \citep{2010PASA...27..379W}.

\subsection{Age and Metallicity Comparisons}

Resolving the questions of UCD connections and origins will require bringing
together more information than just size and luminosity.
Previously, high [$\alpha$/Fe] values measured for UCDs and intermediate-size objects
around various galaxies suggested early, rapid formation, and similarities to normal GCs
rather than to nuclei, which were thought to have more extended star formation histories
\citep{2007A&A...472..111M,2007AJ....133.1722E,2009MNRAS.394.1801F,2010ApJ...712.1191T,2011A&A...525A..86D}.
However, we have seen in Section~\ref{sec:amr} that both M87 UCDs and Virgo nuclei can 
have a range of [$\alpha$/Fe] values, which is further supported by data for
three other UCDs from the literature (plotted in Figure~\ref{fig:amr}, right).
This is also the case for normal GCs 
(e.g., \citealt{2005AJ....130.2140P,2010ApJ...708.1335W}),
so the [$\alpha$/Fe] implications are not yet clear.

Our examination of the age-metallicity relation suggests two types of UCDs:
one type with a similar trend to dE nuclei, and another that is offset to
higher metallicities (Figure~\ref{fig:amr}, left; see also \citealt{2008MNRAS.390..906C}).
In Virgo/M87, the UCD color-magnitude relation tracks that of the nuclei remarkably
well, including a rapid transition to redder colors at high luminosity.
Similarly, \citet{2011ApJ...737...86C} found that UCDs in the Coma cluster include
a population of red objects with similar ages and metallicities to elliptical
galaxies, as well as blue objects more similar to dE nuclei.

Such distinctions are reminiscent of the inner and outer halo GCs of the Milky Way
(e.g., \citealt{2010MNRAS.404.1203F,2011ApJ...738...74D}), and suggest
that there may be generically two populations of UCDs: one associated with {\it massive} (and therefore metal-rich) bulge
formation, and another with dwarf galaxy accretion.
Galaxy-to-galaxy variations in the prevalence of these two populations 
(e.g., NGC~1399 vs.\, M87; \citealt{2006AJ....131.2442M}) might then reveal differences in 
star formation history and galaxy assembly
(cf \citealt{2008ApJ...689..936J}).  It is worth noting that an association with massive bulge formation
could encompass UCDs as remnant nuclei of massive galaxies and/or as products of star cluster collisions
during bulge formation. 

\subsection{Orbital Dynamics}

Another fundamental but relatively unexplored line of evidence for UCD origins is
their orbital dynamics, as encoded in their positions, velocities, and sizes.
We have carried out the first ever detailed analysis of size-distance trends in a UCD system
using our M87 dataset, finding that most of the UCDs show little evidence of following
standard expectations from tidal limitation.  
A comparably weak size-distance trend was also found for luminous UCDs in the core of
Coma \citep{2011ApJ...737...86C}.
Although more theoretical work is needed,
these findings may favor a giant GC formation scenario,
rather than ECs, merged star clusters,
or stripped nuclei which might be expected to show tidal trends
(e.g., \citealt{2011A&A...529A.138B}).
The exception is the subset of low-luminosity M87 UCDs that appear to join up
with some of the more compact objects to define a tidally-limited size-distance trend,
suggesting that they are a family of ECs.

In M87, there is a significant kinematical difference between blue GCs and
blue UCDs (including intermediate-size objects for the sake of useful statistics),
such that the UCDs have a broader distribution of velocities.
This finding strengthens the
conclusion that UCDs are a population that is distinct from compact GCs.
However, EC or stripped nuclei origins are not yet clearly
distinguished by the kinematics, possibly because there is more than one class of UCD
(Section~\ref{sec:kin}).
The conclusions are also limited by the lack of clear theoretical predictions for
kinematics.

In other galaxies, higher velocity dispersions and more flat-topped velocity distributions
have been found for the brighter ``GCs'', 
which is similar to the trends in M87 if the brighter ``GCs'' with unmeasured sizes
are assumed to be mostly UCDs
\citep{2009AJ....137.4956R,2010A&A...513A..52S,2010AJ....139.1871W}.
The ``UCD'' system of NGC 1399, the central massive elliptical in the Fornax cluster,
has been studied in some detail.
Here the UCDs' compact spatial distribution and low projected velocity dispersion 
toward the galaxy center are suggestive of star cluster origins, with further work
still needed to consider trends with color, luminosity, and size
(\citealt{2008MNRAS.389..102T,2009AJ....137..498G};
cf also \citealt{2007ApJ...668L..35W,2011A&A...531A...4M}).

\subsection{Other Clues to be Explored}

More detailed analyses of associations in kinematic/metallicity/size/luminosity/position 
phase-space may provide insight into the origins and interconnections of
UCDs, GCs, nuclei, and ECs.
For instance, UCDs that were formerly galactic nuclei might be accompanied by their original satellite GCs; or UCDs that formed as star clusters from a discrete star-forming event along with other GCs, would be expected to form a phase-space group with their associated star clusters. A related prediction is that in a star cluster scenario, eventually more than one UCD will be found in a phase-space ``group'', while this should never happen in the nuclei scenario. 
In M87, there are indications of phase-space associations (S+11) that motivate more effort in this area. 

The {\it internal} dynamics of UCDs is another standard but challenging approach for
deciphering their origins.  Measurements of velocity dispersion can be used to
map out fundamental-plane type relations with other stellar systems, and to test
for mass-to-light ratio variations, including those due to any remaining dark matter 
accompanying threshed galaxy nuclei
(e.g., \citealt{2004ApJ...610..233M,2005ApJ...627..203H,2007AJ....133.1722E,2009MNRAS.396.1619C,2010ApJ...712.1191T,2011ApJ...726..108T,2011MNRAS.412.1627C,2011MNRAS.413.2665F,2011MNRAS.414L..70F}).
So far, some UCDs have been found to have elevated mass-to-light ratios that may imply
nuclei origins, and some have not.  It is possible that multiple UCD types are again
implied, and more work is needed to connect mass-to-light ratio trends with 
size, luminosity, metallicity, etc.  In particular, velocity dispersion measurements
of the new class of low-luminosity UCDs would be invaluable.

Other properties of UCDs that should also reflect their origins include
their ellipticities and internal color gradients.
However, due to the observational challenges,
these aspects have not yet been examined sufficiently to provide strong evidence for
any particular formational mechanism
(e.g., \citealt{2008AJ....136..461E}).

\subsection{The Stripping Scenario}

\begin{figure}
\vspace{0.1cm}
\epsscale{0.8}
\plotone{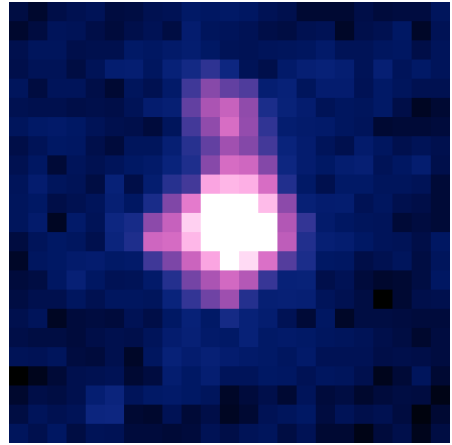}
\figcaption[s923_new3]{\label{fig:weird}
The peculiar object S923, at a projected distance of 3.6~arcmin (17~kpc) from M87,
with the same field-of-view as the sub-panels in Figure~\ref{fig:thumb}.
Its line-of-sight velocity is $2777\pm26$~\kms\ (S+11), which is barely within the standard range for the
Virgo cluster \citep{2002MNRAS.335..712T}, suggesting a highly radial orbit near
its closest approach to M87.
The image is from {\it HST}/WFPC2 in the F606W band;
the object has a core with a size of $\sim$~2~pc, and two asymmetric protuberances
extending out to distances of $\sim$~35--60~pc from the center.
It is fairly faint and blue overall: $M_V=-9.1$, $(g-i)_0=0.71$.
This object might be a dwarf galaxy nucleus caught in the act of stripping.
}
\end{figure}

If nuclei are stripped to form UCDs, then we should occasionally see transition objects,
typically with relatively high luminosities, large sizes, and 
signs of residual stellar material.
Many ``two-component'' objects with extended stellar envelopes have indeed been found,
both around M87 (VUCD7, S7023) and other galaxies
(e.g., \citealt{2003Natur.423..519D,2005ApJ...627..203H,2008MNRAS.385L..83C,2011ApJ...737...86C}).
Other more irregular objects with asymmetric extensions have also been found around M87 (S923;
see Figure~\ref{fig:weird}) and NGC~1399 (78:12; \citealt{2005A&A...439..533R}).

Many nuclei are observed to be compact, and the M87 intermediate-size ($\sim$~5--10~pc) objects have similar kinematical 
and color-magnitude behavior to the UCDs. These facts,
along with the indications in the Local Group that objects like $\omega$~Cen, M54, and G1 
(with $r_{\rm h}\sim$~3--7~pc) are stripped nuclei (e.g., \citealt{2005MNRAS.360..631M}),
all suggest that many intermediate-size objects could be considered as stripped-nucleus UCDs.

The general correspondence of many UCDs with galaxy nuclei in parameter spaces
involving color, luminosity, age, metallicity, $\alpha$-element enhancement, 
and size (after possible evolution) supports a connection between these objects.
Most of the metal-poor and metal-rich UCDs would then originate in the nuclei of dEs and
more massive galaxies, respectively, with a subset of tidally limited faint objects that may
be more closely related to star clusters.  

These interpretations can be connected to a broad framework for different modes
of galaxy assembly (e.g., \citealt{2007Natur.450.1020C,2009ApJ...702.1058Z,2010MNRAS.404.1203F,2011MNRAS.413.2665F,2010ApJ...725.2312O,2011ApJ...736L..26A,2011MNRAS.416.2802F}).
The blue UCDs formed from stripped nuclei 
can be considered part of the outer halo build-up by accretion of low-mass 
infalling satellites
at relatively late epochs, depositing accompanying GCs and field stars as well.  
The red UCDs could be associated with metal-rich bulge formation in the context of
larger galaxies whose nuclei, {\it or satellite extended clusters}, became these UCDs.

\section{Summary and Conclusions}\label{sec:concl}

We have used {\it HST} imaging, and spectroscopy from Keck and the MMT, to double the sample of UCDs around M87. These 34 UCDs form the largest such confirmed sample around any galaxy. Assisted by precise size measurements and deep spectroscopy, we have been able to explore a new region of the size-luminosity parameter space and have discovered objects with large sizes but low luminosities -- a new type of UCD. Their inclusion erases the size--luminosity relation that was formerly thought to exist for, and even define, UCDs. The new sample includes the lowest surface brightness cases confirmed to date.  

This unprecedentedly large, wide-ranging, and homogeneous sample informs our exploration of the origins of these extended star systems that we have attempted to generalize in the context of the ``zoo" of galaxies and star clusters. 
After initially defining UCDs as objects in the $r_{\rm h}\sim$~10--100~pc size range, 
we find the following results of note:
\begin{itemize}
\item The new class of low surface brightness UCD identified in M87 has a few counterparts in other systems, including NGC~2419 in the Milky Way halo.
\item Another $\sim$~50 UCDs can probably be found around M87 through further {\it HST} and spectroscopic surveys.  Special attention is warranted for four candidate low-surface brightness UCDs identified in the central regions.
\item The GC sizes in M87 correlate with luminosity and color but not with galactocentric distance.  The lack of a distance trend argues against the GC sizes being driven by tidal limitations.
\item The UCD sizes in M87 show a weak overall anti-correlation with distance, and no
overall luminosity dependence.  Again, tidal effects are not apparent.
\item A subset of lower luminosity GCs and UCDs near the central regions of M87 appear to show a tight size-distance trend that may imply tidal effects for these objects.
\item The region of size-luminosity parameter space occupied by the M87 UCDs does not match up well with Virgo galaxy nuclei, unless age differences and UCD expansion are invoked.
\item In M87, most of the UCDs are blue and show a tight color-magnitude relation with a small ``tilt'', offset from the trend for the blue GCs.  A small subset of the UCDs extends to red colors and high luminosities.
\item Based on spectroscopic studies from the literature, both in M87 and elsewhere, UCDs are generally old and have a range of $\alpha$-enhancement.  They may split into two subpopulations following tracks in age and metallicity space that are analogous to inner and outer halo GCs in the Milky Way.  It is not clear whether this is a true bimodality, or part of a continuous trend with metallicity or luminosity.
\item The Virgo dE nuclei show remarkable coincidence with M87's blue UCDs in color-magnitude and age-metallicity space.  These similarities may strengthen when only nuclei in denser environments are considered.  The red UCDs may likewise coincide with the nuclei of more massive galaxies.
\item The M87 UCDs have a higher line-of-sight velocity dispersion, and less peaked velocity distribution, than the blue GCs.  There are hints that this effect is driven by the brighter UCDs.
\item M87 objects in the \rh~$\sim$~5--10~pc intermediate-size range show similarities to the larger objects in their kinematics and in the color-magnitude diagram.  It may therefore be appropriate to empirically revise the size boundary between UCDs and bright GCs to $r_{\rm h} \sim$~5~pc, which would encompass objects like $\omega$~Cen.
\item There is a diagonal gap between galaxies and compact stellar systems in size-luminosity space.  The presence of a few bridging objects between UCDs and compact ellipticals suggests that at least some of the more luminous UCDs are related to galaxies.
\item There is a hint of a size gap among the UCDs at $r_{\rm h} \sim$~15--20 pc, which
might imply the existence of UCD subpopulations.
\item There is a zone of avoidance between UCDs and ECs that might be a product of observational selection effects, requiring investigation through more systematic surveys.
\item Two peculiar compact objects are identified that could be UCDs in the process of forming, and that merit additional study.
\end{itemize}

We synthesize this novel set of empirical constraints into
implications for UCD formation in the context of four scenarios:  
giant GCs, merged clusters, luminous ECs, and stripped nuclei.
The giant GC scenario appears to be strongly disfavored as a major contributor
to the UCD population in M87, where we find differences between UCDs
and blue GCs in kinematics and in color-magnitude space, and do not find the expected
strong size-luminosity relation at the bright end.
The color-magnitude finding also appears to argue against a merged GC scenario.

We propose that UCDs, both in M87 and in general,
consist of three generic subpopulations that overlap in size-luminosity
parameter space, requiring the joint study of other parameters to tease out the distinctions.
The majority of the blue UCDs are identified with the threshed nuclei of dwarf ellipticals.
The bright, red UCDs may be associated with the remnants of more massive, metal-rich galaxies.
It is not yet clear if these would represent the actual nuclei (in which case this subpopulation
is contiguous with the blue UCDs), or would be associated with merged clusters.
Some of the less luminous, blue UCDs may be bona-fide star clusters. Whether these are 
equivalent to bright ECs, or have originated by merging, may be settled by determining
whether or not the zone of avoidance is a real gap.

Each of these scenarios requires further theoretical analyses of dynamics and stellar 
populations to see if these can reproduce the observational results on orbital properties and
the relations between size, luminosity, age, and metallicity.
It will also be essential to increase the UCD sample size by pushing to lower luminosities in other galaxies, to consider carefully the observational selection effects,
and to make more detailed studies of ages and metallicities in UCDs and galaxy nuclei.
Such inquiries will provide new insight into both the formation of star clusters and the assembly of galaxy halos.

\acknowledgments

We thank Kristin Chiboucas, S{\o}ren Larsen, Juan Madrid, and Ingo Misgeld for providing their data in electronic form, Holger Baumgardt, Kenji Bekki, Mark Gieles, Sheila Kannappan, Pavel Kroupa and  Mark Norris for helpful conversations, and an anonymous referee for constructive suggestions.  Some of the observations reported here were obtained at the MMT Observatory, a joint facility of the Smithsonian Institution and the University of Arizona. This paper uses data products produced by the OIR Telescope Data Center, supported by the Smithsonian Astrophysical Observatory. Much of the data presented herein were obtained at the W.~M.~Keck Observatory, which is operated as a scientific partnership among the California Institute of Technology, the University of California and the National Aeronautics and Space Administration. Based on observations made with the NASA/ESA Hubble Space Telescope, and obtained from the Hubble Legacy Archive, which is a collaboration between the Space Telescope Science Institute (STScI/NASA), the Space Telescope European Coordinating Facility (ST-ECF/ESA) and the Canadian Astronomy Data Centre (CADC/NRC/CSA). This work was supported by the National Science Foundation through grants AST-0808099, AST-0909237, AST-1101733, and AST-1109878.
We acknowledge financial support from the {\it Access to Major Research Facilities
Programme}, a component of the {\it International Science
Linkages Programme} established under the Australian
Government's innovation statement, {\it Backing Australia's Ability}.

\newpage

\appendix

\section{A. Extended database of sizes and luminosities}\label{sec:uber}

Here we present a new dataset of sizes and luminosities for
stellar systems of a broad range of types in the nearby Universe,
from tiny star clusters and dwarf galaxies, to giant elliptical galaxies.
Many of the datasets assembled in recent years have been oriented around the inclusion of
velocity dispersion estimates
(e.g., \citealt{2008A&A...487..921M,2011ApJ...726..108T},
but focusing on the simple parameters of size and luminosity permits the
assembly of a larger sample that also includes many fainter objects.
Such a database was published by \citet{2011MNRAS.414.3699M},
and serves as one of our major sources of data. We update it as described below,
both by incorporating additional data and by excluding objects with
uncertain properties.

Because of our interest in rare objects with unusual properties, it is
important to include only data with well-constrained measurements.
The basic parameters are the $V$-band absolute magnitude $M_V$, and
the half-light radius \rh\ in physical units, which means that an object's distance
must be fairly well established.  Otherwise, a fuzzy object that appears compact
on the sky might be a relatively nearby GC or a dwarf galaxy, or alternatively a
more distant giant galaxy.  In many cases, distances are established via
spectroscopic redshifts, while the most nearby objects may be recognized because
they can be resolved, at least partially, into individual stars.

A related concern is to avoid objects with large, and potentially uncertain,
degrees of reddening from dust obscuration.  
This applies to the Milky Way and to M31, where we include only those objects with
inferred extinction values of $A_V \la 1$~mag.
Where possible, we also restrict our sample to those objects with overall ages of $\ga$~5~Gyr
in order to minimize the scatter in luminosity that can
result from stellar mass-to-light ratio variations.
This means omitting spiral galaxies and interesting extended objects like Hodge~4 and W3,
while we also exclude sub-components of galaxies like bulges and nuclei.

We begin with the M87 GCs and UCDs presented here and in \citet{Strader11},
and add in objects culled from the literature that meet our criteria.
Two of our largest sources of data are the catalog of Milky Way GCs from
\citet{2010arXiv1012.3224H}, and the compilation of galaxies and star clusters
from \citet{2011MNRAS.414.3699M}.
The other sources are listed in the Table notes; in some cases these are not
the original sources for the measurements, but provide compilations of 
previous data from the literature.

In general, we have not attempted to correct for variations in distance scales
between different studies.
Many of the objects did not have $V$-band magnitudes reported, and we have had
to estimate these through approximate color transformations.
For one dataset \citep{2003AJ....126.1794G}, we converted the tabulated
semi-major axis effective radii to circularized half-light radii based on 
the ellipticities; the dSph catalog of \citet{2011arXiv1106.5500B} ought to be
corrected in the same way, but we do not have the ellipticities available.

In order to avoid biasing the categorization of these objects
(as star clusters, dwarf galaxies, etc.), we have not added any such classifiers,
but only applied some suggestive labels to broad areas of parameter space in
Figure~\ref{fig:uber}.  Very roughly, using preliminary classifications as discussed in
this paper, the extended database of 970 objects includes
$\sim$~400~GCs, $\sim$~100~ECs, $\sim$~50~intermediate objects, $~\sim$~100~UCDs,
$\sim$~50~dSphs, $\sim$~100~dEs and cEs, and $\sim$~100~gEs.

The datatable is also available on the SAGES webpage:
{\tt http://sages.ucolick.org/spectral\_database.html }.

\newpage

\LongTables 

\begin{deluxetable}{lcccc}
\tablewidth{0pt}
\tabletypesize{\footnotesize}
\tablecaption{Catalog of sizes and luminosities for nearby stellar systems \label{tab:uber}}
\tablehead{ID & Host / Environment & $M_V$ & $\log_{10} r_{\rm h}$ & Reference \\
& &  & [pc] & \\ }
\startdata
   Segue~III  & Milky~Way &  $-$0.06 & 0.342 & F+11c \\
SDSS~J1058+2843 & Milky~Way &  $-$0.2  & 1.342 & M+08 \\
  Koposov~2   & Milky~Way &  $-$0.61 & 0.324 & H10 \\
     Segue~I  & Milky~Way &  $-$1.5  & 1.462 & MH11 \\
        AM~4  & Milky~Way &  $-$1.97 & 0.603 & H10 \\
    Segue~II  & Milky~Way &  $-$2.5  & 1.531 & MH11 \\
   Whiting~1  & Milky~Way &  $-$2.56 & 0.285 & H10 \\
 Bo\"otes~II  & Milky~Way &  $-$2.7  & 1.708 & MH11 \\
   Willman~I  & Milky~Way &  $-$2.7  & 1.398 & MH11 \\
       Pal~1  & Milky~Way &  $-$3.01 & 0.169 & H10 \\
       Pal~4  & Milky~Way &  $-$3.14 & 1.207 & H10 \\
      M33-D   &  M33      &  $-$3.79 & 0.681 & C+11a \\
      Pal~13  & Milky~Way &  $-$3.92 & 0.434 & H10 \\
      BH24    &  M31      &  $-$4.01 & 0.379 & B+07 \\
Coma~Berenices & Milky~Way &  $-$4.1 & 1.886 & MH11 \\
Ursa~Major~II & Milky~Way &  $-$4.2  & 2.146 & MH11 \\
  Koposov~1   & Milky~Way &  $-$4.28 & 0.562 & H10 \\
ESO121-SC03   &  LMC      &  $-$4.37 & 1.000 & VM04 \\
      HEC13   &  M31      &  $-$4.4  & 1.431 & H+11a \\
       HEC3   &  M31      &  $-$4.5  & 1.322 & H+11a \\
      Pal~12  & Milky~Way &  $-$4.53 & 0.978 & H10 \\
        AM~1  & Milky~Way &  $-$4.73 & 1.167 & H10 \\
      BH11    &  M31      &  $-$4.82 & 0.341 & B+07 \\
Canes~Venatici~II & Milky~Way &  $-$4.9  & 1.869 & MH11 \\
      Pal~14  & Milky~Way &  $-$4.93 & 1.432 & H10 \\
      BH04    &  M31      &  $-$4.99 & 0.477 & B+07 \\
   Pisces~II  & Milky~Way &  $-$5.0  & 1.778 & MH11 \\
      BH29    &  M31      &  $-$5.03 & 0.518 & B+07 \\
      M33-E   &  M33      &  $-$5.09 & 0.716 & C+11a \\
   ESO-SC06   & Milky~Way &  $-$5.10 & 0.814 & H10 \\
       HEC8   &  M31      &  $-$5.1  & 1.342 & H+11a \\
   Eridanus   & Milky~Way &  $-$5.19 & 1.080 & H10 \\
     GC0606   &  NGC~5128 &  $-$5.2  & 1.230 & M+10 \\
       Leo~V  & Milky~Way &  $-$5.2  & 2.124 & MH11 \\
  Reticulum   &  LMC      &  $-$5.22 & 1.286 & VM04 \\
  NGC~6822~C3 & NGC~6822 &  $-$5.22 & 0.875 & H+11b \\
   Terzan~7   & Milky~Way &  $-$5.24 & 0.707 & H10 \\
       Pal~5  & Milky~Way &  $-$5.27 & 1.264 & H10 \\
       HEC2   &  M31      &  $-$5.3  & 1.230 & H+11a \\
     DAO38    &  M31      &  $-$5.43 & 0.399 & B+07 \\
   Fornax~1   &  Fornax   &  $-$5.44 & 1.094 & MH11 \\
   Terzan~8   & Milky~Way &  $-$5.46 & 0.858 & H10 \\
Ursa~Major~I  & Milky~Way &  $-$5.5  & 2.502 & MH11 \\
       HEC6   &  M31      &  $-$5.5  & 1.380 & H+11a \\
      M091    &  M31      &  $-$5.54 & 0.550 & B+07 \\
       Arp~2  & Milky~Way &  $-$5.61 & 1.165 & H10 \\
      Leo~IV  & Milky~Way &  $-$5.8  & 2.314 & MH11 \\
   NGC~7492   & Milky~Way &  $-$5.81 & 0.944 & H10 \\
       Pal~3  & Milky~Way &  $-$5.82 & 1.241 & H10 \\
      B333    &  M31      &  $-$5.84 & 0.467 & B+07 \\
     B056D    &  M31      &  $-$5.89 & 0.337 & B+07 \\
      M33-A   &  M33      &  $-$5.89 & 1.308 & C+11a \\
       HEC9   &  M31      &  $-$6.0  & 1.380 & H+11a \\
       HEC1   &  M31      &  $-$6.0  & 1.380 & H+11a \\
  NGC~6822~C4 & NGC~6822 &  $-$6.06 & 1.140 & H+11b \\
  NGC~6822~C2 & NGC~6822 &  $-$6.10 & 1.061 & H+11b \\
      B279    &  M31      &  $-$6.13 & 0.328 & B+07 \\
      M33-S   &  M33      &  $-$6.19 & 1.223 & C+11a \\
...\\
\enddata
\tablecomments{
Only a portion of this table is provided for the sake of illustration.
The full table will be provided with the published paper and at:
{\tt http://sages.ucolick.org/spectral\_database.html }.
The data table is arranged in ascending order by luminosity.
All magnitudes are extinction corrected.
In many cases the precision of the size or magnitude measurements is
higher than the actual measurement uncertainties, but the additional
significant figures are kept in order to avoid discreteness effects in the plots.
The reference abbreviations are as follows:
B+07: \citet{2007AJ....133.2764B};
B+11: \citet{2011arXiv1106.5500B};
BL02: \citet{2002AJ....124.1410B};
C+09: \citet{2009Sci...326.1379C};
C+11a: \citet{2011ApJ...730..112C};
C+11b: \citet{2011ApJ...737...86C};
D+09: \citet{2009AJ....137.4361D};
F+11a: \citet{2011MNRAS.413.2665F};
F+11b: \citet{2011MNRAS.415.3393F};
F+11c: \citet{2011AJ....142...88F};
G+03: \citet{2003AJ....126.1794G};
H+05: \citet{2005ApJ...627..203H};
H07: \citet{2007PhDT.........4H};
H+09: \citet{2009MNRAS.394L..97H};
H10: \citet{2010arXiv1012.3224H};
H+11a: \citet{2011MNRAS.414..770H};
H+11b: \citet{2011ApJ...738...58H};
LB00: \citet{2000AJ....120.2938L};
M+07: \citet{2007A&A...472..111M};
M+08: \citet{2008ApJ...684.1075M};
M+10: \citet{2010MNRAS.404.1157M};
M+11: \citet{2011A&A...531A...4M};
MH04: \citet{2004ApJ...610..233M};
MH11: \citet{2011MNRAS.414.3699M};
NK11: \citet{2011MNRAS.414..739N};
R+05: \citet{2005A&A...439..533R};
R+07: \citet{2007A&A...469..147R};
R+09: \citet{2009AJ....137.4956R};
S+08: \citet{2008MNRAS.391..685S};
S+11: \citet{Strader11};
VM04: \citet{2004MNRAS.354..713V}.
}
\end{deluxetable}


\begin{thebibliography}{}

\bibitem[Arnold et al.(2011)]{2011ApJ...736L..26A} Arnold, J.~A., Romanowsky, A.~J., Brodie, J.~P., Chomiuk, L., Spitler, L.~R., Strader, J., Benson, A.~J., \& Forbes, D.~A.\ 2011, \apjl, 736, L26 
\bibitem[Bailin \& Harris(2009)]{2009ApJ...695.1082B} Bailin, J., \& Harris, W.~E.\ 2009, \apj, 695, 1082 
\bibitem[Barmby et al.(2007)]{2007AJ....133.2764B} Barmby, P., McLaughlin, D.~E., Harris, W.~E., Harris, G.~L.~H., \& Forbes, D.~A.\ 2007, \aj, 133, 2764
\bibitem[Bassino et al.(1994)]{1994ApJ...431..634B} Bassino, L.~P., Muzzio, J.~C., \& Rabolli, M.\ 1994, \apj, 431, 634 
\bibitem[Baumgardt \& Mieske(2008)]{2008MNRAS.391..942B} Baumgardt, H., \& Mieske, S.\ 2008, \mnras, 391, 942 
\bibitem[Baumgardt et al.(2010)]{2010MNRAS.401.1832B} Baumgardt, H., Parmentier, G., Gieles, M., \& Vesperini, E.\ 2010, \mnras, 401, 1832 
\bibitem[Bekki et al.(2001)]{2001ApJ...552L.105B} Bekki, K., Couch, W.~J., \& Drinkwater, M.~J.\ 2001, \apjl, 552, L105 
\bibitem[Bekki et al.(2003)]{2003MNRAS.344..399B} Bekki, K., Couch, W.~J., Drinkwater, M.~J., \& Shioya, Y.\ 2003, \mnras, 344, 399
\bibitem[Bekki(2007)]{2007MNRAS.380.1177B} Bekki, K.\ 2007, \mnras, 380, 1177
\bibitem[Blakeslee \& Barber DeGraaff(2008)]{2008AJ....136.2295B} Blakeslee, J.~P., \& Barber DeGraaff, R.\ 2008, \aj, 136, 2295 
\bibitem[Brasseur et al.(2011)]{2011arXiv1106.5500B} Brasseur, C.~M., Martin, N.~F., Macci{\`o}, A.~V., Rix, H.-W., \& Kang, X.\ 2011, \apj, submitted, arXiv:1106.5500 
\bibitem[Brodie \& Larsen(2002)]{2002AJ....124.1410B} Brodie, J.~P., \& Larsen, S.~S.\ 2002, \aj, 124, 1410 
\bibitem[Br{\"u}ns et al.(2009)]{2009ApJ...702.1268B} Br{\"u}ns, R.~C., Kroupa, P., \& Fellhauer, M.\ 2009, \apj, 702, 1268 
\bibitem[Br{\"u}ns et al.(2011)]{2011A&A...529A.138B} Br{\"u}ns, R.~C., Kroupa, P., Fellhauer, M., Metz, M., \& Assmann, P.\ 2011, \aap, 529, A138 
\bibitem[Burkert et al.(2005)]{2005ApJ...628..231B} Burkert, A., Brodie, J., \& Larsen, S.\ 2005, \apj, 628, 231 
\bibitem[Cantiello et al.(2007)]{2007ApJ...668..209C} Cantiello, M., Blakeslee, J.~P., \& Raimondo, G.\ 2007, \apj, 668, 209 
\bibitem[Carlson \& Holtzman(2001)]{2001PASP..113.1522C} Carlson, M.~N., \& Holtzman, J.~A.\ 2001, \pasp, 113, 1522 
\bibitem[Carollo et al.(2007)]{2007Natur.450.1020C} Carollo, D., et al.\ 2007, \nat, 450, 1020
\bibitem[Chiboucas et al.(2011)]{2011ApJ...737...86C} Chiboucas, K., Tully, R.~B., Marzke, R.~O., et al.\ 2011, \apj, 737, 86 
\bibitem[Chies-Santos et al.(2007)]{2007A&A...467.1003C} Chies-Santos, A.~L., Santiago, B.~X., \& Pastoriza, M.~G.\ 2007, \aap, 467, 1003 
\bibitem[Chilingarian \& Mamon(2008)]{2008MNRAS.385L..83C} Chilingarian, I.~V., \& Mamon, G.~A.\ 2008, \mnras, 385, L83 
\bibitem[Chilingarian et al.(2008)]{2008MNRAS.390..906C} Chilingarian, I.~V., Cayatte, V., \& Bergond, G.\ 2008, \mnras, 390, 906 
\bibitem[Chilingarian et al.(2009)]{2009Sci...326.1379C} Chilingarian, I., Cayatte, V., Revaz, Y., Dodonov, S., Durand, D., Durret, F., Micol, A., \& Slezak, E.\ 2009, Science, 326, 1379 
\bibitem[Chilingarian et al.(2011)]{2011MNRAS.412.1627C} Chilingarian, I.~V., Mieske, S., Hilker, M., \& Infante, L.\ 2011, \mnras, 412, 1627 
\bibitem[Cockcroft et al.(2011)]{2011ApJ...730..112C} Cockcroft, R., et al.\ 2011, \apj, 730, 112 
\bibitem[Cohen \& Ryzhov(1997)]{1997ApJ...486..230C} Cohen, J.~G., \& Ryzhov, A.\ 1997, \apj, 486, 230
\bibitem[Cohen et al.(2010)]{2010ApJ...725..288C} Cohen, J.~G., Kirby, E.~N., Simon, J.~D., \& Geha, M.\ 2010, \apj, 725, 288 
\bibitem[Collins et al.(2009)]{2009MNRAS.396.1619C} Collins, M.~L.~M., et al.\ 2009, \mnras, 396, 1619 
al.\ 2009, \mnras, 396, 1619 
\bibitem[C{\^o}t{\'e} et al.(2006)]{2006ApJS..165...57C} C{\^o}t{\'e}, P., et al.\ 2006, \apjs, 165, 57 
\bibitem[Dabringhausen et al.(2008)]{2008MNRAS.386..864D} Dabringhausen, J., Hilker, M., \& Kroupa, P.\ 2008, \mnras, 386, 864 
\bibitem[Da Costa et al.(2009)]{2009AJ....137.4361D} Da Costa, G.~S., Grebel, E.~K., Jerjen, H., Rejkuba, M., \& Sharina, M.~E.\ 2009, \aj, 137, 4361 
\bibitem[Da Rocha et al.(2011)]{2011A&A...525A..86D} Da Rocha, C., Mieske, S., Georgiev, I.~Y., Hilker, M., Ziegler, B.~L., \& Mendes de Oliveira, C.\ 2011, \aap, 525, A86 
\bibitem[Dotter et al.(2011)]{2011ApJ...738...74D} Dotter, A., Sarajedini, A., \& Anderson, J.\ 2011, \apj, 738, 74 
\bibitem[Drinkwater et al.(2000)]{2000PASA...17..227D} Drinkwater, M.~J., Jones, J.~B., Gregg, M.~D., \& Phillipps, S.\ 2000, PASA, 17, 227 
\bibitem[Drinkwater et al.(2003)]{2003Natur.423..519D} Drinkwater, M.~J., Gregg, M.~D., Hilker, M., Bekki, K., Couch, W.~J., Ferguson, H.~C., Jones, J.~B., \& Phillipps, S.\ 2003, \nat, 423, 519 
\bibitem[Elmegreen(2008)]{2008ApJ...672.1006E} Elmegreen, B.~G.\ 2008, \apj, 672, 1006 
\bibitem[Evstigneeva et al.(2007)]{2007AJ....133.1722E} Evstigneeva, E.~A., Gregg, M.~D., Drinkwater, M.~J., \& Hilker, M.\ 2007, \aj, 133, 1722 
\bibitem[Evstigneeva et al.(2008)]{2008AJ....136..461E} Evstigneeva, E.~A., et al.\ 2008, \aj, 136, 461 
\bibitem[Faber(1973)]{1973ApJ...179..423F} Faber, S.~M.\ 1973, \apj, 179, 423 
\bibitem[Fadely et al.(2011)]{2011AJ....142...88F} Fadely, R., Willman, B., Geha, M., et al.\ 2011, \aj, 142, 88 
\bibitem[Fellhauer \& Kroupa(2002)]{2002MNRAS.330..642F} Fellhauer, M., \& Kroupa, P.\ 2002, \mnras, 330, 642 
\bibitem[Fellhauer \& Kroupa(2005)]{2005MNRAS.359..223F} Fellhauer, M., \& Kroupa, P.\ 2005, \mnras, 359, 223 
\bibitem[Firth et al.(2007)]{2007MNRAS.382.1342F} Firth, P., Drinkwater, M.~J., Evstigneeva, E.~A., et al.\ 2007, \mnras, 382, 1342 
\bibitem[Firth et al.(2008)]{2008MNRAS.389.1539F} Firth, P., Drinkwater, M.~J., \& Karick, A.~M.\ 2008, \mnras, 389, 1539 
\bibitem[Firth et al.(2009)]{2009MNRAS.394.1801F} Firth, P., Evstigneeva, E.~A., \& Drinkwater, M.~J.\ 2009, \mnras, 394, 1801 
\bibitem[Font et al.(2011)]{2011MNRAS.416.2802F} Font, A.~S., McCarthy, I.~G., Crain, R.~A., et al.\ 2011, \mnras, 416, 2802
\bibitem[Forbes et al.(2008)]{2008MNRAS.389.1924F} Forbes, D.~A., Lasky, P., Graham, A.~W., \& Spitler, L.\ 2008, \mnras, 389, 1924 
\bibitem[Forbes \& Bridges(2010)]{2010MNRAS.404.1203F} Forbes, D.~A., \& Bridges, T.\ 2010, \mnras, 404, 1203
\bibitem[Forbes \& Kroupa(2011)]{2011PASA...28...77F} Forbes, D.~A., \& Kroupa, P.\ 2011, PASA, 28, 77 
\bibitem[Forbes et al.(2011)]{2011MNRAS.413.2665F} Forbes, D.~A., Spitler, L.~R., Graham, A.~W., Foster, C., Hau, G.~K.~T., \& Benson, A.\ 2011, \mnras, 413, 2665 
\bibitem[Foster et al.(2011)]{2011MNRAS.415.3393F} Foster, C., Spitler, L.~R., Romanowsky, A.~J., et al.\ 2011, \mnras, 415, 3393 
\bibitem[Frank et al.(2011)]{2011MNRAS.414L..70F} Frank, M.~J., Hilker, M., Mieske, S., Baumgardt, H., Grebel, E.~K., \& Infante, L.\ 2011, \mnras, 414, L70 
\bibitem[Freeman(1993)]{1993ASPC...48..608F} Freeman, K.~C.\ 1993, The Globular Cluster-Galaxy Connection, 48, 608 
\bibitem[Geha et al.(2003)]{2003AJ....126.1794G} Geha, M., Guhathakurta, P., \& van der Marel, R.~P.\ 2003, \aj, 126, 1794 
\bibitem[Gieles et al.(2010)]{2010MNRAS.408L..16G} Gieles, M., Baumgardt, H., Heggie, D.~C., \& Lamers, H.~J.~G.~L.~M.\ 2010, \mnras, 408, L16 
\bibitem[Gieles et al.(2011)]{2011MNRAS.413.2509G} Gieles, M., Heggie, D.~C., \& Zhao, H.\ 2011, \mnras, 413, 2509 
\bibitem[Gilmore et al.(2007)]{2007ApJ...663..948G} Gilmore, G., Wilkinson, M.~I., Wyse, R.~F.~G., Kleyna, J.~T., Koch, A., Evans, N.~W., \& Grebel, E.~K.\ 2007, \apj, 663, 948 
\bibitem[Goerdt et al.(2008)]{2008MNRAS.385.2136G} Goerdt, T., Moore, B., Kazantzidis, S., Kaufmann, T., Macci{\`o}, A.~V., \& Stadel, J.\ 2008, \mnras, 385, 2136
\bibitem[Gregg et al.(2009)]{2009AJ....137..498G} Gregg, M.~D., et al.\ 2009, \aj, 137, 498 
\bibitem[Ha{\c s}egan et al.(2005)]{2005ApJ...627..203H} Ha{\c s}egan, M., et al.\ 2005, \apj, 627, 203 
\bibitem[Ha{\c s}egan(2007)]{2007PhDT.........4H} Ha{\c s}egan, I.~M.\ 2007, Ph.D.~Thesis, Rutgers Univ. 
\bibitem[Hanes et al.(2001)]{2001ApJ...559..812H} Hanes, D.~A., C{\^o}t{\'e}, P., Bridges, T.~J., McLaughlin, D.~E., Geisler, D., Harris, G.~L.~H., Hesser, J.~E., \& Lee, M.~G.\ 2001, \apj, 559, 812 
\bibitem[Harris et al.(2006)]{2006ApJ...636...90H} Harris, W.~E., Whitmore, B.~C., Karakla, D., Oko{\'n}, W., Baum, W.~A., Hanes, D.~A., \& Kavelaars, J.~J.\ 2006, \apj, 636, 90 
\bibitem[Harris(2009a)]{2009ApJ...699..254H} Harris, W.~E.\ 2009a, \apj, 699, 254 
\bibitem[Harris(2009b)]{2009ApJ...703..939H} Harris, W.~E.\ 2009b, \apj, 703, 939 
\bibitem[Harris(2010)]{2010arXiv1012.3224H} Harris, W.~E.\ 2010, arXiv:1012.3224
\bibitem[Hau et al.(2009)]{2009MNRAS.394L..97H} Hau, G.~K.~T., Spitler, L.~R., Forbes, D.~A., Proctor, R.~N., Strader, J., Mendel, J.~T., Brodie, J.~P., \& Harris, W.~E.\ 2009, \mnras, 394, L97 
\bibitem[Hilker et al.(1999)]{1999A&AS..134...75H} Hilker, M., Infante, L., Vieira, G., Kissler-Patig, M., \& Richtler, T.\ 1999, \aaps, 134, 75 
\bibitem[Hilker(2009)]{2009gcgg.book...51H} Hilker, M.\ 2009, in Globular Clusters - Guides to Galaxies, 51 
\bibitem[Hopkins et al.(2010)]{2010MNRAS.401L..19H} Hopkins, P.~F., Murray, N., Quataert, E., \& Thompson, T.~A.\ 2010, \mnras, 401, L19 
\bibitem[Huchra \& Brodie(1984)]{1984ApJ...280..547H} Huchra, J., \& Brodie, J.\ 1984, \apj, 280, 547 
\bibitem[Huchra \& Brodie(1987)]{1987AJ.....93..779H} Huchra, J., \& Brodie, J.\ 1987, \aj, 93, 779
\bibitem[Hurley \& Mackey(2010)]{2010MNRAS.408.2353H} Hurley, J.~R., \& Mackey, A.~D.\ 2010, \mnras, 408, 2353 
\bibitem[Huxor et al.(2005)]{2005MNRAS.360.1007H} Huxor, A.~P., Tanvir, N.~R., Irwin, M.~J., Ibata, R., Collett, J.~L., Ferguson, A.~M.~N., Bridges, T., \& Lewis, G.~F.\ 2005, \mnras, 360, 1007 
\bibitem[Huxor et al.(2008)]{2008MNRAS.385.1989H} Huxor, A.~P., Tanvir, N.~R., Ferguson, A.~M.~N., Irwin, M.~J., Ibata, R., Bridges, T., \& Lewis, G.~F.\ 2008, \mnras, 385, 1989 
\bibitem[Huxor et al.(2009)]{2009ApJ...698L..77H} Huxor, A., Ferguson, A.~M.~N., Barker, M.~K., Tanvir, N.~R., Irwin, M.~J., Chapman, S.~C., Ibata, R., \& Lewis, G.\ 2009, \apjl, 698, L77 
\bibitem[Huxor et al.(2011a)]{2011MNRAS.414..770H} Huxor, A.~P., et al.\ 2011a, \mnras, 414, 770
\bibitem[Huxor et al.(2011b)]{2011MNRAS.414.3557H} Huxor, A.~P., Phillipps, S., Price, J., \& Harniman, R.\ 2011b, \mnras, 414, 3557 
\bibitem[Hwang \& Lee(2006)]{2006ApJ...638L..79H} Hwang, N., \& Lee, M.~G.\ 2006, \apjl, 638, L79 
\bibitem[Hwang \& Lee(2008)]{2008AJ....135.1567H} Hwang, N., \& Lee, M.~G.\ 2008, \aj, 135, 1567 
\bibitem[Hwang et al.(2011)]{2011ApJ...738...58H} Hwang, N., Lee, M.~G., Lee, J.~C., et al.\ 2011, \apj, 738, 58 
\bibitem[Johnston et al.(2008)]{2008ApJ...689..936J} Johnston, K.~V., Bullock, J.~S., Sharma, S., Font, A., Robertson, B.~E., \& Leitner, S.~N.\ 2008, \apj, 689, 936 
\bibitem[Jones et al.(2006)]{2006AJ....131..312J} Jones, J.~B., et al.\ 2006, \aj, 131, 312 
\bibitem[Jord{\'a}n et al.(2005)]{2005ApJ...634.1002J} Jord{\'a}n, A., et al.\ 2005, \apj, 634, 1002 
\bibitem[Jord{\'a}n et al.(2009)]{2009ApJS..180...54J} Jord{\'a}n, A., et al.\ 2009, \apjs, 180, 54 
\bibitem[Kissler-Patig et al.(2006)]{2006A&A...448.1031K} Kissler-Patig, M., Jord{\'a}n, A., \& Bastian, N.\ 2006, \aap, 448, 1031 
\bibitem[Kroupa(1998)]{1998MNRAS.300..200K} Kroupa, P.\ 1998, \mnras, 300, 200 
\bibitem[K{\"u}pper et al.(2008)]{2008MNRAS.389..889K} K{\"u}pper, A.~H.~W., Kroupa, P., \& Baumgardt, H.\ 2008, \mnras, 389, 889 
\bibitem[Larsen(1999)]{1999A&AS..139..393L} Larsen, S.~S.\ 1999, \aaps, 139, 393 
\bibitem[Larsen \& Brodie(2000)]{2000AJ....120.2938L} Larsen, S.~S., \& Brodie, J.~P.\ 2000, \aj, 120, 2938 
\bibitem[Larsen \& Brodie(2002)]{2002AJ....123.1488L} Larsen, S.~S., \& Brodie, J.~P.\ 2002, \aj, 123, 1488 
\bibitem[Larsen \& Brodie(2003)]{2003ApJ...593..340L} Larsen, S.~S., \& Brodie, J.~P.\ 2003, \apj, 593, 340 
\bibitem[Lotz et al.(2004)]{2004ApJ...613..262L} Lotz, J.~M., Miller, B.~W., \& Ferguson, H.~C.\ 2004, \apj, 613, 262 
\bibitem[Mackey \& van den Bergh(2005)]{2005MNRAS.360..631M} Mackey, A.~D., \& van den Bergh, S.\ 2005, \mnras, 360, 631 
\bibitem[Mackey et al.(2006)]{2006ApJ...653L.105M} Mackey, A.~D., et al.\ 2006, \apjl, 653, L105 
\bibitem[Madrid et al.(2009)]{2009ApJ...705..237M} Madrid, J.~P., Harris, W.~E., Blakeslee, J.~P., \& G{\'o}mez, M.\ 2009, \apj, 705, 237 
\bibitem[Madrid et al.(2010)]{2010ApJ...722.1707M} Madrid, J.~P., et al.\ 2010, \apj, 722, 1707 
\bibitem[Madrid(2011)]{2011ApJ...737L..13M} Madrid, J.~P.\ 2011, \apjl, 737, L13 
\bibitem[Maraston et al.(2004)]{2004A&A...416..467M} Maraston, C., Bastian, N., Saglia, R.~P., et al.\ 2004, \aap, 416, 467 
\bibitem[Martin et al.(2008)]{2008ApJ...684.1075M} Martin, N.~F., de Jong, J.~T.~A., \& Rix, H.-W.\ 2008, \apj, 684, 1075 
\bibitem[Martini \& Ho(2004)]{2004ApJ...610..233M} Martini, P., \& Ho, L.~C.\ 2004, \apj, 610, 233 
\bibitem[Masters et al.(2010)]{2010ApJ...715.1419M} Masters, K.~L., et al.\ 2010, \apj, 715, 1419
\bibitem[McLaughlin(2000)]{2000ApJ...539..618M} McLaughlin, D.~E.\ 2000, \apj, 539, 618 
\bibitem[Mieske et al.(2006a)]{2006AJ....131.2442M} Mieske, S., Hilker, M., Infante, L., \& Jord{\'a}n, A.\ 2006a, \aj, 131, 2442 
\bibitem[Mieske et al.(2006b)]{2006ApJ...653..193M} Mieske, S., et al.\ 2006b, \apj, 653, 193 
\bibitem[Mieske et al.(2007)]{2007A&A...472..111M} Mieske, S., Hilker, M., Jord{\'a}n, A., Infante, L., \& Kissler-Patig, M.\ 2007, \aap, 472, 111 
\bibitem[Mieske et al.(2008)]{2008A&A...487..921M} Mieske, S., et al.\ 2008, \aap, 487, 921 
\bibitem[Misgeld et al.(2011)]{2011A&A...531A...4M} Misgeld, I., Mieske, S., Hilker, M., Richtler, T., Georgiev, I.~Y., \& Schuberth, Y.\ 2011, \aap, 531, A4 
\bibitem[Misgeld \& Hilker(2011)]{2011MNRAS.414.3699M} Misgeld, I., \& Hilker, M.\ 2011, \mnras, 414, 3699 
\bibitem[Mouhcine et al.(2010)]{2010MNRAS.404.1157M} Mouhcine, M., Harris, W.~E., Ibata, R., \& Rejkuba, M.\ 2010, \mnras, 404, 1157 
\bibitem[Mould et al.(1987)]{1987AJ.....93...53M} Mould, J.~R., Oke, J.~B., \& Nemec, J.~M.\ 1987, \aj, 93, 53
\bibitem[Murray(2009)]{2009ApJ...691..946M} Murray, N.\ 2009, \apj, 691, 946 
\bibitem[Norris \& Kannappan(2011)]{2011MNRAS.414..739N} Norris, M.~A., \& Kannappan, S.~J.\ 2011, \mnras, 414, 739 
\bibitem[Oser et al.(2010)]{2010ApJ...725.2312O} Oser, L., Ostriker, J.~P., Naab, T., Johansson, P.~H., \& Burkert, A.\ 2010, \apj, 725, 2312
\bibitem[Paolillo et al.(2011)]{2011ApJ...736...90P} Paolillo, M., Puzia, T.~H., Goudfrooij, P., et al.\ 2011, \apj, 736, 90 
\bibitem[Paudel et al.(2010)]{2010ApJ...724L..64P} Paudel, S., Lisker, T., \& Janz, J.\ 2010, \apjl, 724, L64 
\bibitem[Paudel et al.(2011)]{2011MNRAS.413.1764P} Paudel, S., Lisker, T., \& Kuntschner, H.\ 2011, \mnras, 413, 1764 
\bibitem[Peng et al.(2006)]{2006ApJ...639..838P} Peng, E.~W., et al.\ 2006, \apj, 639, 838
\bibitem[Pfalzner(2009)]{2009A&A...498L..37P} Pfalzner, S.\ 2009, \aap, 498, L37 
\bibitem[Phillipps et al.(2001)]{2001ApJ...560..201P} Phillipps, S., Drinkwater, M.~J., Gregg, M.~D., \& Jones, J.~B.\ 2001, \apj, 560, 201 
\bibitem[Pritzl et al.(2005)]{2005AJ....130.2140P} Pritzl, B.~J., Venn, K.~A., \& Irwin, M.\ 2005, \aj, 130, 2140 
\bibitem[Rejkuba et al.(2007)]{2007A&A...469..147R} Rejkuba, M., Dubath, P., Minniti, D., \& Meylan, G.\ 2007, \aap, 469, 147 
\bibitem[Richardson \& Green(1997)]{RichGreen} Richardson, S., \& Green, P.\ 1997, J.~R. Statistical Soc.~B, 59, 731
\bibitem[Richtler et al.(2005)]{2005A&A...439..533R} Richtler, T., Dirsch, B., Larsen, S., Hilker, M., \& Infante, L.\ 2005, \aap, 439, 533 
\bibitem[Richtler(2006)]{2006BASI...34...83R} Richtler, T.\ 2006, Bulletin of the Astronomical Society of India, 34, 83 
\bibitem[Romanowsky et al.(2009)]{2009AJ....137.4956R} Romanowsky, A.~J., Strader, J., Spitler, L.~R., Johnson, R., Brodie, J.~P., Forbes, D.~A., \& Ponman, T.\ 2009, \aj, 137, 4956
\bibitem[Scheepmaker et al.(2007)]{2007A&A...469..925S} Scheepmaker, R.~A., Haas, M.~R., Gieles, M., Bastian, N., Larsen, S.~S., \& Lamers, H.~J.~G.~L.~M.\ 2007, \aap, 469, 925 
\bibitem[Schuberth et al.(2010)]{2010A&A...513A..52S} Schuberth, Y., Richtler, T., Hilker, M., Dirsch, B., Bassino, L.~P., Romanowsky, A.~J., \& Infante, L.\ 2010, \aap, 513, A52
\bibitem[Sharina et al.(2005)]{2005A&A...442...85S} Sharina, M.~E., Puzia, T.~H., \& Makarov, D.~I.\ 2005, \aap, 442, 85 
\bibitem[Sinnott et al.(2010)]{2010AJ....140.2101S} Sinnott, B., Hou, A., Anderson, R., Harris, W.~E., \& Woodley, K.~A.\ 2010, \aj, 140, 2101
\bibitem[Smith Castelli et al.(2008)]{2008MNRAS.391..685S} Smith Castelli, A.~V., Faifer, F.~R., Richtler, T., \& Bassino, L.~P.\ 2008, \mnras, 391, 685 
\bibitem[Spitler et al.(2006)]{2006AJ....132.1593S} Spitler, L.~R., Larsen, S.~S., Strader, J., Brodie, J.~P., Forbes, D.~A., \& Beasley, M.~A.\ 2006, \aj, 132, 1593 
\bibitem[Stonkut{\.e} et al.(2008)]{2008AJ....135.1482S} Stonkut{\.e}, R., et al.\ 2008, \aj, 135, 1482 
\bibitem[Strader et al.(2006)]{2006AJ....132.2333S} Strader, J., Brodie, J.~P., Spitler, L., \& Beasley, M.~A.\ 2006, \aj, 132, 2333
\bibitem[Strader \& Smith(2008)]{2008AJ....136.1828S} Strader, J., \& Smith, G.~H.\ 2008, \aj, 136, 1828 
\bibitem[Strader et al.(2011)]{Strader11} Strader, J., et al.\ 2011, \apjs, submitted (S+11)
\bibitem[Taylor et al.(2010)]{2010ApJ...712.1191T} Taylor, M.~A., Puzia, T.~H., Harris, G.~L., Harris, W.~E., Kissler-Patig, M., \& Hilker, M.\ 2010, \apj, 712, 1191 
\bibitem[Thomas et al.(2008)]{2008MNRAS.389..102T} Thomas, P.~A., Drinkwater, M.~J., \& Evstigneeva, E.\ 2008, \mnras, 389, 102 
\bibitem[Tollerud et al.(2011)]{2011ApJ...726..108T} Tollerud, E.~J., Bullock, J.~S., Graves, G.~J., \& Wolf, J.\ 2011, \apj, 726, 108 
\bibitem[Tolstoy et al.(2009)]{2009ARA&A..47..371T} Tolstoy, E., Hill, V., \& Tosi, M.\ 2009, \araa, 47, 371 
\bibitem[Trentham \& Tully(2002)]{2002MNRAS.335..712T} Trentham, N., \& Tully, R.~B.\ 2002, \mnras, 335, 712 
\bibitem[van den Bergh et al.(1991)]{1991ApJ...375..594V} van den Bergh, S., Morbey, C., \& Pazder, J.\ 1991, \apj, 375, 594 
\bibitem[van den Bergh(1995)]{1995AJ....110.1171V} van den Bergh, S.\ 1995, \aj, 110, 1171 
\bibitem[van den Bergh \& Mackey(2004)]{2004MNRAS.354..713V} van den Bergh, S., \& Mackey, A.~D.\ 2004, \mnras, 354, 713 
\bibitem[van den Bergh(2008)]{2008MNRAS.390L..51V} van den Bergh, S.\ 2008, \mnras, 390, L51 
\bibitem[Wehner \& Harris(2007)]{2007ApJ...668L..35W} Wehner, E.~M.~H., \& Harris, W.~E.\ 2007, \apjl, 668, L35 
\bibitem[Woodley et al.(2010a)]{2010ApJ...708.1335W} Woodley, K.~A., Harris, W.~E., Puzia, T.~H., G{\'o}mez, M., Harris, G.~L.~H., \& Geisler, D.\ 2010a, \apj, 708, 1335 
\bibitem[Woodley et al.(2010b)]{2010AJ....139.1871W} Woodley, K.~A., G{\'o}mez, M., Harris, W.~E., Geisler, D., \& Harris, G.~L.~H.\ 2010b, \aj, 139, 1871
\bibitem[Woodley \& G{\'o}mez(2010)]{2010PASA...27..379W} Woodley, K.~A., \& G{\'o}mez, M.\ 2010, PASA, 27, 379 
\bibitem[Zolotov et al.(2009)]{2009ApJ...702.1058Z} Zolotov, A., Willman, B., Brooks, A.~M., Governato, F., Brook, C.~B., Hogg, D.~W., Quinn, T., \& Stinson, G.\ 2009, \apj, 702, 1058

\end{thebibliography}
\end{document}